\documentclass[prd,onecolumn,showpacs,11pt]{revtex4-1}
\usepackage[ngerman]{babel}
\usepackage[utf8]{inputenc}
\usepackage{amsmath, amssymb, amsfonts}
\usepackage{natbib}
\usepackage{hyperref}
\usepackage{color}
\hypersetup{colorlinks=true}

\begin{document}
\title{Holography as a highly efficient RG flow I : Rephrasing gravity}
%\preprint{}
\author{Nicolas Behr}
\email{nicolas.behr@ed.ac.uk}
\affiliation{LFCS, University of Edinburgh, Informatics Forum, 10 Crichton Street, Edinburgh, EH8 9AB, Scotland, UK}
\affiliation{Maxwell Institute for Mathematical Sciences, Edinburgh, UK}
\author{Stanislav Kuperstein}
\email{stanislav.kuperstein@gmail.com}
\affiliation{Institut de Physique Th\'{e}orique, CEA Saclay, F-91191 Gif-sur-Yvette, France}
\author{Ayan Mukhopadhyay}
\email{ayan@physics.uoc.gr}
\affiliation{Centre de Physique Th\'{e}orique, Ecole Polytechnique, CNRS, 91128 Palaiseau Cedex, France}
\affiliation{Institut de Physique Th\'{e}orique, CEA Saclay, F-91191 Gif-sur-Yvette, France}
\affiliation{Crete Center for Theoretical Physics, University of Crete, 71003 Heraklion, Greece \footnote{Present address of A.M. since October 1, 2014}}
%\emailAdd{Nicolas.Behr@gmx.de}
%\emailAdd{stanislav.kuperstein@gmail.com}
%\emailAdd{ayan@physics.uoc.gr}
%\preprint{CCTP-2014-26, CCQCN-2014-51}
%%%%% for reference: old title page:

%
%\author{$\text{Nicolas Behr,}^{a,b}$
%$\text{Brian P. Dolan,}^{a,b}$
%$\text{Stanislav Kuperstein,}^{c}$
%$\text{and Ayan Mukhopadhyay}^{c,d}$\\
%
%% The "\note" macro will give a warning: "Ignoring empty anchor..."
%% you can safely ignore it.
%\textsf{a} \textit{Department of Mathematics, Heriot-Watt University, Riccarton, Edinburgh, EH14 4AS, U.K.}\\
%\textsf{b} \textit{Maxwell Institute for Mathematical Sciences, Edinburgh, U.K.}\\
%\textsf{c} \textit{Institut de Physique Th\'{e}orique, CEA Saclay, F-91191 Gif-sur-Yvette, France}\\
%\textsf{d} \textit{Centre de Physique Th\'{e}orique, Ecole Polytechnique, 91128 Palaiseau Cedex, France}\\
%
%
%% e-mail addresses: one for each author, in the same order as the authors
%% 
%\texttt{emails}:
%\textsf{N.Behr@hw.ac.uk, b.p.dolan@hw.ac.uk, stanislav.kuperstein@cea.fr, mukhopadhyay@cpht.polytechnique.fr}}
%

\begin{abstract}
We investigate how the holographic correspondence can be reformulated as a generalisation of Wilsonian RG flow in a strongly interacting large $N$ quantum field theory. We firstly define a \textit{highly efficient RG flow} as one in which the Ward identities related to local conservation of energy, momentum and charges preserve the same form at each scale -- to achieve this it is necessary to redefine the background metric and external sources at each scale as functionals of the effective single trace operators. These redefinitions also absorb the contributions of the multi-trace operators to these effective Ward identities. Thus the background metric and external sources become effectively dynamical reproducing the dual classical gravity equations in one higher dimension. Here, we focus on reconstructing the pure gravity sector as a highly efficient RG flow of the energy-momentum tensor operator, leaving the explicit constructive field theory approach for generating such RG flows to the second part of the work. We show that special symmetries of the highly efficient RG flows carry information through which we can decode the gauge fixing of bulk diffeomorphisms in the corresponding gravity equations. We also show that the highly efficient RG flow which reproduces a given classical gravity theory in a given gauge is \textit{unique} provided the endpoint can be transformed to a non-relativistic fixed point with a finite number of parameters under a universal rescaling. The results obtained here are used in the second part of this work, where we do an explicit field-theoretic construction of the RG flow, and obtain the dual classical gravity theory.
\end{abstract}
\pacs{11.25.Tq, 11.10.Gh, 04.20.Dw}
\maketitle

\tableofcontents
\section{Introduction}

The holographic correspondence implies that many features of a large class of strongly interacting many--body quantum systems can be understood via a dual classical gravity 
theory with appropriate asymptotic symmetries in one higher dimensional spacetime. Furthermore, the dual classical gravity theory typically should involve only a finite number of fields. 

Since the duality has been first conjectured for the $\mathcal{N}=4$ supersymmetric Yang--Mills (SYM) theory \cite{Maldacena:1997re}, a large body of supporting evidence has been gathered, albeit in a limited number of examples, most of which can be embedded into string theory. It is a widespread belief among experts that the correspondence should be more general. This is supported by the expectation that the emergent extra radial dimension in the dual classical gravity theory encodes a special kind of renormalisation group (RG) flow in a large class of strongly interacting quantum many-body systems. Recently, the works \cite{Heemskerk:2010hk,Faulkner:2010jy,Balasubramanian:2012hb,Lee:2013dln} have stimulated great interest in understanding how the holographic correspondence can be reformulated as a RG flow. 

In this work, which we present in two parts, we show how an explicit constructive field theory approach at strong coupling and large $N$ can reproduce the traditional holographic correspondence \cite{Gubser:1998bc,Witten:1998qj} at least in the dynamical sector corresponding to pure gravity in one higher dimension as a precise kind of \textit{generalisation of Wilsonian RG flow}. 

It turns out that in order to reformulate gravity as a RG flow, it is useful to first rephrase the equations of gravity in a RG-flow-like language, which involves \textit{first order} scale-evolution of physical quantities in a \textit{fixed} non-dynamical background metric. The non-trivial aspect of this rephrasing is that not only should the gravity equations be rewritten in first order form akin to RG flow equations, but also the latter should encode \textit{complete} information of the bulk metric, including the ultraviolet (UV) data that is necessary to produce geometries that have \textit{no naked singularities}. In the first part of this work, which is the subject of the present paper, we show how the equations of $(d+1)-$gravity can be rewritten as such a first order RG flow in a \textit{fixed} non-dynamical background metric, and how the dual classical gravity equations can in turn be reproduced from this RG flow equation. 

The basic goal of the present paper is to find which properties a RG flow in a $d-$dimensional field theory (that defines effective operators at various scales) should have such that the first order scale-evolution equations for operators can be rewritten as equations of a $(d+1)-$ classical theory of gravity with full \textit{diffeomorphism} invariance. We will also determine how such properties of RG flow lead to $(d+1)-$dimensional dual classical geometries which have regular horizons and obey diffeomorphism invariant equations of motion. In the second part of this work \cite{secondpart}, we will give an actual construction of such a RG flow as a precise kind of coarse-graining of the UV dynamics. Thus based on the results of the present paper, we will be able to define \textit{constructive field theory} in the large $N$ and strong coupling limit.

We find that the \textit{key} property of the RG flow which is necessary to obtain diffeomorphism invariance in $(d+1)-$dimensions is that there should exist a metric $g_{\mu\nu}(\Lambda)[{t^\mu_{\phantom{\mu}\nu}}(\Lambda), O_i(\Lambda), \Lambda]$ and background sources $J_i(\Lambda)[{t^\mu_{\phantom{\mu}\nu}}(\Lambda), O_i(\Lambda), \Lambda]$ which are functionals of the single-trace operators at each $\Lambda$ such that the Ward identity for local conservation of energy and momentum can be written in the form 
\begin{equation}\label{WIintro}
\nabla_{(\Lambda)\mu}t^\mu_{\phantom{\mu}\nu}(\Lambda) = \sum_i O_i(\Lambda)\nabla_{(\Lambda)\nu}J_i(\Lambda),
\end{equation}
where $\nabla_{(\Lambda)}$ is the covariant derivative constructed from the effective metric $g_{\mu\nu}(\Lambda)$. Thus an effective dynamical metric and sources emerge out of our construction in order to preserve the form of the Ward identity for local energy-momentum conservation at each scale by absorbing the contributions coming from the multi-trace operators, although the scale evolution of the operators are first order and taking place in the fixed Minkowski space background. Our claim is that the effective metric and sources then leads to a bulk metric and other bulk fields which are solutions of a pure classical gravity theory with \textit{full} diffeomorphism invariance with the identification $r=\Lambda^{-1}$ between the scale $\Lambda$ and the emergent radial coordinate $r$. This feature of preservation of the effective form of the Ward identity will be the key defining feature of what we call a \textit{highly efficient RG flow}. Furthermore, we will see that the map between a highly efficient RG flow and its dual gravity theory is unique.

Independently of reconstruction of holography, we can motivate \textit{highly efficient RG flow} as an appropriate construction for understanding time-dependence of field-theoretic observables. In this case, the directly measurable quantities are collective variables (like hydrodynamic variables, condensates, relaxation modes, etc.) and their statistical fluctuations and correlations. In a strongly coupled large $N$ theory, as will be explicitly demonstrated in \cite{secondpart}, we can define quantum operators at any scale as a functionals of the collective variables which take the same form in each quantum state being measured. The equation of motion for the collective variables lead us to describe the time-dependence in expectation values of the quantum operators. Coarse-graining of collective variables will lead us to construct coarse-grained quantum operators in a state-independent way. This can work only if we preserve the form of Ward identities like that of local energy and momentum conservation, which is violated by projecting out high energy modes that give rise to driving forces on the soft sector. To preserve the Ward identity for the energy-momentum tensor projected onto the soft sector, we need to redefine the effective background metric and sources -- this should be done in such a manner that these are state-independent functionals of the coarse-grained operators (and hence also the collective variables) at each scale. As a result, the directly measurable collective variables will also preserve the form of their equation of motion, although the parameters (like transport coefficients in the hydrodynamic limit) will get renormalised with the scale. In the low energy limit, the dynamics will be captured by a finite number of parameters. It is remarkable that such a structure of the RG flow which follows from the key property (\ref{WIintro}) leads to the emergence of a diffeomorphism-invariant classical gravity theory in one higher dimension, the complete construction of this RG flow will be the objective of the second part \cite{secondpart} of our work.

To proceed further, it is useful to recapitulate some aspects of Wilsonian RG flow. The chief purpose of a RG flow construction is to define effective operators and observables at a given scale of observation. As long as the theory is weakly coupled, it is possible to define all (gauge-invariant) composite operators in terms of effective elementary fields and the effective couplings, like the effective Yang-Mills gauge fields $A_\mu(\Lambda)$ and the effective strong coupling $\alpha_s(\Lambda)$ in pure Quantum Chromodynamics (QCD). For instance, from the Wilsonian effective action $S[A_\mu(\Lambda),\alpha_s(\Lambda), \Lambda]$ of pure QCD at a scale $\Lambda$, we can readily obtain the composite operator $t^\mu_{\phantom{\mu}\nu}(\Lambda)[A_\mu(\Lambda),\alpha_s(\Lambda), \Lambda]$ corresponding to the conserved energy-momentum tensor. By knowing how the elementary fields $A_\mu(\Lambda)$ and the effective coupling(s) $\alpha_s(\Lambda)$ \textit{run} with the scale, we can obtain the scale-evolution equations of composite operators of the form:
\begin{equation}\label{op-evol}
\frac{\partial}{\partial\Lambda} \mathcal{O}_m(\Lambda) = \sum_{n} C_{mn}(\Lambda) \mathcal{O}_n(\Lambda).
\end{equation}
If the theory is asymptotically free, we can relate operators at a given scale to those of the free theory in the UV, thus leading to a precise definition of the effective operator algebra at a given scale.

In the case of a strongly interacting large $N$ conformal field theory (CFT), we can expect that only a few single-trace operators will have small scaling dimensions, and the rest will decouple from the RG flow (\ref{op-evol}) since the coefficients will be suppressed by large powers of $\Lambda^{-1}$. Nevertheless, as has been emphasised in \cite{Heemskerk:2010hk}, the multi-trace operators built out of the surviving single-trace operators will also appear in (\ref{op-evol}). In the large $N$ limit, the expectation values of the multi-trace operators will factorise in \textit{all} states. Therefore, we can regard (\ref{op-evol}) as non-linear \textit{classical} equations for the scale-evolution of the single-trace operators -- the notion of classicality is applied here in the sense that we can replace the multi-trace operators as the product of the expectation values of the respective single-trace operators in  (\ref{op-evol}). Thus, in this paper we will drop our caution in replacing operators by their expectation values, and not make any distinction between them, although we will almost always mean the latter.

Furthermore, we will assume that there exists a subspace in the Hilbert space where the operator $t^\mu_{\phantom{\mu}\nu}$ is the \textit{only} single-trace operator that has a non-trivial expectation value, viz. the expectation value of other single-trace operators either vanish, or are the same for all states in this sector. If the holographic correspondence is indeed reproduced, this assumption is justified as we can almost always consistently truncate the dual classical gravity theory to pure gravity. In our approach, we start with this assumption, however we should justify this assumption along with those assumptions mentioned in the previous paragraph \textit{a posteriori} by establishing a consistent definition of an effective operator algebra at all scales. In this sector of states, (\ref{op-evol}) reduces to:
\begin{equation}\label{t-evol-intro}
\frac{\partial}{\partial \Lambda}t^\mu_{\phantom{\mu}\nu}(\Lambda) = F^\mu_{\phantom{\mu}\nu}[t^\mu_{\phantom{\mu}\nu}(\Lambda),\Lambda ],
\end{equation}
Since in a CFT, we have no other intrinsic scale, the above equation should have an expansion of the form:
\begin{eqnarray}\label{t-evol-exp}
\frac{\partial}{\partial \Lambda}t^\mu_{\phantom{\mu}\nu}(\Lambda) &=& a_{3}\left(\lambda_{(i)}\right)\cdot \frac{1}{\Lambda^3}\cdot \Box {t^\mu_{\phantom{\mu}\nu}}(\Lambda) +
\nonumber\\&&
+  \frac{1}{\Lambda^5}\cdot \Big(a_{5a}\left(\lambda_{(i)}\right)\cdot  {t^\mu_{\phantom{\mu}\rho}}(\Lambda)
{t^\rho_{\phantom{\rho}\nu}}(\Lambda) +a_{5b}\left(\lambda_{(i)}\right)\cdot \delta^\mu_{\phantom{\mu}\nu} \cdot {t^\alpha_{\phantom{\alpha}\beta}}(\Lambda)
{t^\beta_{\phantom{\beta}\alpha}}(\Lambda)+ \nonumber\\ &&\quad \quad\quad+a_{5c}\left(\lambda_{(i)}\right)\cdot \Box^2 {t^\mu_{\phantom{\mu}\nu}}(\Lambda)\Big)+\cdots\,,
\end{eqnarray}
where $\lambda_{(i)}$ are the non-running coupling constants of the CFT. We have also assumed that only $t^\mu_{\phantom{\mu}\nu}$ and multi-trace operators built out of it can appear in the above equation as mentioned before. The above expansion is valid for sufficiently large $\Lambda$ satisfying $\Lambda \gg \Lambda_{\rm IR}$, where $\Lambda_{\rm IR}$ is a state-dependent scale related to the expectation value of ${t^\mu_{\phantom{\mu}\nu}}^\infty$, which is defined as the microscopic UV operator, i.e. ${t^\mu_{\phantom{\mu}\nu}}(\Lambda=\infty)$. Above we have used the CFT Ward identities $\partial_\mu {t^\mu_{\phantom{\mu}\nu}}^\infty = 0$ and ${\rm Tr}\, {t}^\infty = 0$ to constrain (\ref{t-evol-exp}) up to redundancies. In the second part of our work \cite{secondpart}, we will also extend our construction partly by including another scalar single-trace operator.

%Although we will be generalising Wilsonian RG flow, we will still aim to reconstruct pure gravity as a first order RG flow of the form (\ref{t-evol-intro}) or equivalently (\ref{t-evol-exp}) in a \textit{fixed} background metric, which for most of this paper will be assumed to be flat Minkowski space. The key difference will be that unlike the Wilsonian RG flow, (i) we will use collective variables instead of elementary fields that parametrise the expectation values to construct the operator ${t^\mu_{\phantom{\mu}\nu}}(\Lambda)$, and (ii) we will promote the cut-off in momentum space to a functional of these collective variables -- the precise details will appear in the second part of this work \cite{secondpart}. 

We repeat once again that we will leave the actual field theory construction of (\ref{t-evol-intro}) to the second part of this work. In the present paper, we will focus on how the equations of pure classical gravity can be rewritten into the form (\ref{t-evol-intro}) and vice versa. Furthermore, we will see in the second part of this work that there is a \textit{unique} RG flow equation (\ref{t-evol-intro}) corresponding to a given dual theory of gravity such that it has an endpoint at a finite scale $\Lambda_{\rm IR}$ (where the effective metric $g_{\mu\nu}(\Lambda)$ becomes non-invertible and various components of $t^\mu_{\phantom{\mu}\nu}(\Lambda)$ blow up) which can be transformed to a non-relativistic \textit{fixed} point with a finite number of parameters under the universal rescaling:
\begin{equation}\label{rescale-intro}
\frac{1}{\Lambda_{\rm IR}} - \frac{1}{\Lambda} = \frac{\xi}{\tilde{\Lambda}}, \quad t = \frac{\tau}{\xi}, \quad \text{and $\xi \rightarrow 0$ with $\tilde{\Lambda}, \tau$ held fixed.}
\end{equation}
Furthermore, the transformation of the endpoint to the non-relativistic fixed point will be possible only if the UV data is chosen uniquely -- otherwise various components of $t^\mu_{\phantom{\mu}\nu}(\Lambda)$ will blow up at $\Lambda = \Lambda_{\rm IR}$ beyond prescribed bounds that should be satisfied in order to ensure the existence of the non-relativistic fixed point after the rescaling (\ref{rescale-intro}). Indeed the prescribed bounds determine the integration constants of the first order RG flow and thus also determine the UV data. Remarkably, these UV data are precisely those which are necessary for ensuring that the dual geometries on the gravity side do not have naked singularities, matching those which are traditionally calculated by solving the classical gravity equations explicitly. Therefore, our RG flow construction of the form (\ref{t-evol-intro}) will be able to also reproduce the UV data that corresponds to regular solutions of the dual gravity theory. 

In this paper, we will also discuss briefly how our construction of the RG flow in the form (\ref{t-evol-exp}) will ensure the good infrared behaviour discussed above. Nevertheless, in order to see this explicitly, we will need to sum over all orders in $\Lambda^{-1}$, which will be explicitly possible only in some dynamical limits. We will return to this in the second part of our work. 

We will call our RG flows \textit{highly efficient RG flows} because they are able to preserve the effective form of the Ward identity for local conservation of energy and momentum at each scale by redefining the background, and also because they are able to self determine the UV data. In this way, we will be able to somewhat mimic what Wilsonian RG flow achieves in an asymptotically free theory, namely defining the effective observables at a scale in terms of the ultraviolet free theory, although here we will achieve the same via the \textit{hidden} infrared non-relativistic fixed point which has a finite number of parameters. For instance, in the hydrodynamic limit the latter will be the incompressible non-relativistic Navier-Stokes equations.

In this paper, we will also see how generalisations of (\ref{t-evol-intro}), where other auxiliary non-dynamical state-dependent variables can appear, will lead to classical gravity equations in arbitrary gauges, which need not be the Fefferman-Graham gauge. In all cases, the effective Ward identity (\ref{WIintro}) will be preserved at each scale. Furthermore, we will show that one can read off the choice of gauge fixing on the gravity side from the symmetries of the RG flow equations (\ref{t-evol-intro}) when the latter are constructed on a conformally flat background metric. In particular, these symmetries of the RG flow will be automorphic to the residual gauge symmetries on the gravity side that map to conformal transformations on the boundary. We will call such symmetries of the RG flows which reveal the geometry of hypersurface foliations on the gravity side, in which effective observables are evaluated, the \textit{lifted Weyl symmetries}. Via these results, we will be able to achieve complete translations between pure classical gravity equations and highly efficient RG flows for any arbitrary gauge-fixing of bulk diffeomorphisms.

Although our work is inspired by \cite{Heemskerk:2010hk,Faulkner:2010jy,Lee:2013dln}, we differ in many key points. The major point of departure is that we consider a generalisation of Wilsonian RG flow in order to reproduce gravity. Also unlike \cite{Lee:2013dln}, we do not project the Wilsonian RG flow to the space of single-trace operators. Rather, we generalise Wilsonian RG flow by making the cut-off a functional of the collective variables which parametrise the expectation values of single-trace operators, in a manner such that the multi-trace contributions to the Ward identities for local conservation of charge, energy and momentum can be absorbed into new effective backgrounds, so that these retain the same form at each scale. The detailed field-theoretic aspects of our construction will appear in the second part of our work \cite{secondpart}.

The organisation of this paper is as follows: In Section \ref{Intro2}, we will introduce key elements of the definition of highly efficient RG flow, and summarise various conceptual issues which we will be dealing with in this paper and also in the second part of this work. We will also argue why highly efficient RG flows will always lead to classical gravity equations with full diffeomorphism invariance. In Section \ref{explicit-map}, we will show how gravity equations in Fefferman-Graham gauge can be translated to highly efficient RG flows and vice versa. In Section \ref{lifted-Weyl-symmetry}, we will show how bulk diffeomorphisms can be captured in highly efficient RG flows and how the lifted Weyl symmetry reveals the dual geometry of hypersurface foliations. Finally, in Section \ref{conclusions}, we will conclude with a discussion on  how our approach can be extended beyond the pure gravity sector of holography. The appendices provide detailed derivation of many key results of Section \ref{lifted-Weyl-symmetry}.

\section{Introducing the concept of the highly efficient RG flow and how it leads to emergence of gravity}\label{Intro2}

Our central proposition is as follows. Let us consider the $d-$dimensional scale evolution of $t^\mu_{\phantom{\mu}\nu}(\Lambda)$ of the following form:
\begin{equation}\label{schematic1}
\frac{\partial}{\partial \Lambda}t^\mu_{\phantom{\mu}\nu}(\Lambda) = F^\mu_{\phantom{\mu} \nu}[t^\mu_{\phantom{\mu}\nu}(\Lambda),\Lambda ],
\end{equation}
in the \textit{fixed} background metric $g^{\rm (b)}_{\mu\nu}$, such that there exists a background metric $g_{\mu\nu}(\Lambda)$ which is a functional of $t^\mu_{\phantom{\mu}\nu}(\Lambda)$ and $\Lambda$ in the same \textit{fixed} background metric $g^{\rm (b)}_{\mu\nu}$ taking the form
\begin{equation}\label{schematic2}
g_{\mu\nu}(\Lambda) = G_{\mu\nu}[t^\mu_{\phantom{\mu}\nu}(\Lambda),\Lambda ],
\end{equation}
at each $\Lambda$, and in which $t^\mu_{\phantom{\mu}\nu}(\Lambda)$ satisfies the local conservation equation
\begin{equation}\label{WILambda}
\nabla_{(\Lambda)\mu} t^\mu_{\phantom{\mu}\nu}(\Lambda) = 0
\end{equation}
with $\nabla_{(\Lambda)}$ being the covariant derivative constructed from $g_{\mu\nu}(\Lambda)$. Note that $g_{\mu\nu}(\Lambda)$ has to coincide at $\Lambda = \infty$ with the fixed background metric $g^{\rm (b)}_{\mu\nu}$ in which the functionals $F^\mu_{\phantom{\mu} \nu}$ and $G_{\mu\nu}$ are constructed, because ${t^\mu_{\phantom{\mu}\nu}}^\infty$ should satisfy $\nabla_{\rm(b)\mu}{t^\mu_{\phantom{\mu}\nu}}^\infty = 0$, with $\nabla_{\rm (b)}$ being the covariant derivative constructed from $g^{\rm (b)}_{\mu\nu}$. We claim that it follows that $g_{\mu\nu}(\Lambda)$ then gives a bulk metric in the Fefferman-Graham gauge:
\begin{equation}\label{FGmetric}
{\rm d}s^2 = \frac{l^2}{r^2}\left({\rm d}r^2 + g_{\mu\nu}(r,x) {\rm d}x^2\right),
\end{equation}
which solves the equations of a \textit{pure} $(d+1)-$classical gravity theory with \textit{full} $(d+1)-$diffeomorphism invariance and a negative cosmological constant determined by the asymptotic curvature radius $l$, and with $r$ identified with $\Lambda^{-1}$ (i.e. $r = \Lambda^{-1}$). 

This will be the key to recasting a $(d+1)-$diffeomorphism invariant \textit{pure} classical gravity with a negative cosmological constant into the form of a \textit{first order} RG flow and vice versa. A RG flow of the form (\ref{schematic1}) with the above property will be referred to as a \textit{highly efficient RG flow} for the rest of this paper. 

At this point, it may be useful to provide an example. Let us consider the following RG flow equation in $4$ spacetime dimensions:
\begin{eqnarray}\label{t-rg-example}
\frac{\partial t^\mu_{\phantom{\mu}\nu}(\Lambda)}{\partial \Lambda} &=& \frac{1}{\Lambda^3}\cdot\frac{1}{2} \Box t^{\mu}_{\phantom{\mu}\nu}(\Lambda)- \frac{1}{\Lambda^5}\cdot\left(\frac{1}{4}\, \delta^\mu_{\phantom{\mu}\nu}
{t^\alpha_{\phantom{\alpha}\beta}}(\Lambda)
{t^\beta_{\phantom{\beta}\alpha}}(\Lambda) - \frac{7}{128}\,\Box^2 {t^\mu_{\phantom{\mu}\nu}}(\Lambda)\right)-
\nonumber\\&&+\frac{1}{\Lambda^5}\, \log\, \Lambda \cdot \frac{1}{32}\cdot \Box^2 {t^\mu_{\phantom{\mu}\nu}}(\Lambda)
+\mathcal{O}\left(\frac{1}{\Lambda^7}\, \log\, \Lambda\right),
\end{eqnarray}
in flat Minkowski space $\eta_{\mu\nu}$. For the above RG flow, we can indeed construct the following  $g_{\mu\nu}(\Lambda)$:
\begin{eqnarray}\label{g-example}
g_{\mu\nu}(\Lambda) &=& \eta_{\mu\nu} +\, \frac{1}{\Lambda^4}\cdot\frac{1}{4} \eta_{\mu\alpha}{t^\alpha_{\phantom{\alpha}\nu}}(\Lambda)
+\,\frac{1}{\Lambda^6}\cdot \frac{1}{24}\eta_{\mu\alpha}\Box {t^\alpha_{\phantom{\alpha}\nu}}(\Lambda)+
\nonumber\\&&
+ \frac{1}{\Lambda^8} \cdot \left(\frac{1}{32}\,\eta_{\mu\alpha} {t^\alpha_{\phantom{\alpha}\rho}}(\Lambda)
{t^\rho_{\phantom{\rho}\nu}}(\Lambda) -\frac{7}{384}\, \eta_{\mu\nu}{t^\alpha_{\phantom{\alpha}\beta}}(\Lambda)
{t^\beta_{\phantom{\beta}\alpha}}(\Lambda) +\frac{11}{1536}\,\eta_{\mu\alpha}\Box^2 {t^\alpha_{\phantom{\alpha}\nu}}(\Lambda)\right)-
\nonumber\\&& + \frac{1}{\Lambda^8}\,\log \, \Lambda\cdot \frac{1}{516} \cdot \eta_{\mu\alpha}\Box^2 {t^\alpha_{\phantom{\alpha}\nu}}(\Lambda)+ \mathcal{O}\left(\frac{1}{\Lambda^{10}}\, \log\, \Lambda\right), 
\end{eqnarray}
as a functional of $t^\mu_{\phantom{\mu}\nu}(\Lambda)$ and $\Lambda$ in the flat Minkowski space background at each $\Lambda$ such that, when it is considered as a effective background metric, the scale-dependent Ward identity (\ref{WILambda}) is satisfied (given that $\partial_\mu {t^\mu_{\phantom{\mu}\nu}}^\infty = 0)$. This $g_{\mu\nu}(\Lambda)$ is unique up to an overall normalisation of $t^\mu_{\phantom{\mu}\nu}(\Lambda)$ by a ($\Lambda-$indpendent) numerical constant. Furthermore, the $5-$dimensional bulk metric (\ref{FGmetric}) then satisfies Einstein's equations with the cosmological constant set to $-6/l^2$, and with $r = \Lambda^{-1}$. The derivation of the above equations will be given in the next section. The \textit{log} term in (\ref{t-rg-example}) will be shown to be related to the conformal anomaly.

It is to be noted that the Ward identity (\ref{WILambda}) can be recast as an effective operator equation. For example, in the above case (\ref{WILambda}) reduces to:
\begin{eqnarray}
\partial_\mu t^\mu_{\phantom{\mu}\nu}(\Lambda) &=& \frac{1}{\Lambda^4}\cdot\left(\frac{1}{16}\partial_\nu \left(t^\alpha_{\phantom{\alpha}\beta}(\Lambda)t^\beta_{\phantom{\beta}\alpha}(\Lambda)\right)-\frac{1}{8}t^\mu_{\phantom{\mu}\nu}(\Lambda)\partial_\mu\, {\rm Tr}\,t(\Lambda) \right)+\nonumber\\&&
+\frac{1}{\Lambda^6}\cdot\left(\frac{1}{48}t^\alpha_{\phantom{\alpha}\beta}(\Lambda)\partial_\nu\Box t^\beta_{\phantom{\beta}\alpha}(\Lambda)-\frac{1}{48}t^\mu_{\phantom{\mu}\nu}(\Lambda)\partial_\mu\Box\, {\rm Tr}\,t(\Lambda) \right)
+\mathcal{O}\left(\frac{1}{\Lambda^8}\right).
\end{eqnarray}
We can now see that the usual Ward identity is broken at a finite scale by contributions due to multi-trace operators built from $t^\mu_{\phantom{\mu}\nu}(\Lambda)$. Therefore, the scale-dependent effective background $g_{\mu\nu}(\Lambda)$ as given by (\ref{g-example}) serves to absorb these multi-trace contributions in a manner such that the effective Ward identity preserves its form (\ref{WILambda}) at each scale.

At this point, it might seem that the highly efficient RG flow (\ref{schematic1}) with the property that it gives rise to an effective Ward identity of the form (\ref{WILambda}) at each scale is just a strange way to rephrase gravity. However, as we will show in the second part of our work, this is much deeper than just a mere rewriting of classical gravity equations. The other important feature of (\ref{schematic1}) can be seen as follows. We can ask if there exist different functionals $F^\mu_{\phantom{\mu} \nu}$ in (\ref{schematic1}) and also different associated functionals $G_{\mu\nu}$ in (\ref{schematic2}) such that the effective background metric $g_{\mu\nu}(\Lambda)$ remains the same, thus giving rise to the \textit{same} bulk metric (\ref{FGmetric}) that follows the \textit{same} $(d+1)-$classical gravity equations. The answer is that it is indeed the case -- the different choices of $F^\mu_{\phantom{\mu} \nu}$ and $G_{\mu\nu}$ are related to (infinite) ambiguities in gravitational counterterms which define $t^\mu_{\phantom{\mu}\nu}(\Lambda)$ on a hypersurface foliation of spacetime in the dual geometry, and a numerical constant that determines the overall normalisation of $t^\mu_{\phantom{\mu}\nu}(\Lambda)$, as will be shown in the next section. The overall normalisation of $t^\mu_{\phantom{\mu}\nu}(\Lambda)$ can be readily fixed via the two-point correlation function, which can also be obtained from our approach.

We will show in the second part of our work that it follows from the results in \cite{Kuperstein:2013hqa} that only for \textit{unique} choices of functionals $F^\mu_{\phantom{\mu} \nu}$ and $G_{\mu\nu}$ (up to an overall normalisation of $t^\mu_{\phantom{\mu}\nu}$ by a numerical constant), which correspond to \textit{unique} gravitational counterterms on the gravity side, the RG flow has an endpoint at a finite scale $\Lambda_{\rm IR}$ (where the effective metric $g_{\mu\nu}(\Lambda)$ becomes non-invertible and most effective observables blow up) which can be transformed to a non-relativistic fixed point with a \textit{finite} number of parameters under the \textit{universal} rescaling:
\begin{equation}\label{rescaling}
\frac{1}{\Lambda_{\rm IR}} - \frac{1}{\Lambda} = \frac{\xi}{\tilde{\Lambda}}, \quad t = \frac{\tau}{\xi}, \quad \text{and $\xi \rightarrow 0$ with $\tilde{\Lambda}, \tau$ held fixed.}
\end{equation}
Furthermore, the transformation to the non-relativistic fixed point, which in the case of the hydrodynamic limit is given by the non-relativistic incompressible Navier-Stokes equations, is possible \textit{only} if the effective observables such as the transport coefficients satisfy appropriate bounds near $\Lambda = \Lambda_{\rm IR}$ \cite{Kuperstein:2013hqa}. These bounds give the required integration constants of the first order RG flow and thus enable us to determine all the UV data like transport coefficients. Remarkably, these are precisely those values which are required to ensure that the corresponding $(d+1)-$spacetime metric has no naked singularity at the late-time horizon -- based on \cite{Kuperstein:2013hqa} we can verify this explicitly for the case of Einstein's gravity. Thus the highly efficient RG flow equation (\ref{schematic1}) with a unique choice of $F^\mu_{\phantom{\mu} \nu}$ corresponding to a given dual pure gravity theory can also determine the UV data which lead to absence of naked singularities in the emergent spacetime, and which are traditionally obtained by solving the dual classical gravity equations explicitly. We will return to this aspect in detail in the second part of this work.

Our proposition for rewriting classical gravity theory as a highly efficient RG flow works also when we include higher derivative corrections to Einstein's gravity, provided these corrections are treated perturbatively. We will discuss this aspect in further detail in the next section. Nevertheless, it is not clear whether for arbitrary higher derivative corrections to Einstein's gravity we can claim that the UV data that ensures regularity of the solutions of gravity also implies that the end point of a highly efficient RG flow can be transformed to a non-relativistic fixed point with finite number of parameters after the rescaling (\ref{rescaling}), as our calculations will be presented explicitly for the case of Einstein's gravity. We conjecture that such a gravity theory, where the criterion of absence of naked singularity cannot be translated into good infrared behaviour of a RG flow, does not correspond to any consistent dual quantum field theory. We will return to this issue again in the second part of our work.

Another immediate question that arises is that if there is anything special about the choice of the Fefferman-Graham gauge that may be necessary to recast the classical gravity equations as highly efficient RG flows. Indeed we will see in Section \ref{lifted-Weyl-symmetry} that it is also possible to write the gravity equations in any other gauge in the form of a highly efficient RG flow and vice versa, but the scale-evolution equation (\ref{schematic1}) then also contains auxiliary (non-dynamical) variables, which are related to the lapse function and shift vector in the bulk metric. Furthermore, the symmetries of the RG flow equations, which are related to residual gauge symmetries on the gravity side that map to conformal transformations in the UV, contain direct information about the corresponding gauge fixing of the $(d+1)-$diffeomorphisms, and thus also about the foliation of hypersurfaces in the dual $(d+1)-$geometry where $t^\mu_{\phantom{\mu}\nu}(\Lambda)$ and the effective metric $g_{\mu\nu}(\Lambda)$ are evaluated.

\paragraph*{Why highly efficient RG flows should reproduce $(d+1)-$diffeomorphism invariance:} Before we go into the details of the construction of the highly efficient RG flow, let us consider why our definition of the highly efficient RG flow ensures that we reproduce a classical gravity theory with full diffeomorphism invariance. Firstly, it is clear that if a theory of gravity does not have full diffeomorphism invariance, it should have extra degrees of freedom -- particularly the lapse function and shift vectors will be dynamical as is the case in Horava-Lifshitz gravity. In presence of such extra degrees of freedom, we expect to reproduce a Ward identity at best of the form:
\begin{equation}
\nabla_\mu t^\mu_{\phantom{\mu}\nu} = \text{terms involving $O_i$ and $J_i$}
\end{equation}
where the $O_i$ are the single-trace operators dual to the extra propagating bulk fields, and where the $J_i$ are the corresponding external sources. Therefore, \textit{it will be impossible to truncate the RG flow consistently such that the only single-trace operator that appears all through the scale evolution is the energy-momentum tensor}. We also cannot satisfy the criterion of the highly efficient RG flow which demands that the effective Ward identity should always take the form of (\ref{WILambda}) at \textit{all} scales. Indeed the Ward identities that follow from holographic analysis of Horava-Lifshitz gravity have been obtained explicitly in \cite{Chemissany:2014xsa} (see also \cite{Ross:2011gu,Hartong:2014oma}), and these do not take the form given by (\ref{WILambda}). This can be furthermore verified by the fact that in the hydrodynamic limit, the holographic analysis of Horava-Lifshitz gravity does not reproduce standard Navier-Stokes equations, but those in presence of additional sources (torsion and Newton potential) \cite{Kiritsis:2015doa}. In any case, we should prove more rigorously in the future that it is possible to construct an effective Ward identity of the form (\ref{WILambda}) in a RG flow \textit{only} if the dual gravity theory has full $(d+1)-$diffeomorphism invariance. 

On the other hand, if we indeed have a classical gravity theory with full $(d+1)-$diffeomorphism invariance, it is always possible to construct a $t^\mu_{\phantom{\mu}\nu}(\Lambda)$ in the corresponding RG flow equations that satisfies the effective Ward identity (\ref{WILambda}) at all scales by virtue of the fact that the classical gravity equations satisfy $(d+1)-$Bianchi identities. The explicit demonstration of this assertion will be given in the following section.

\section{Gravity equations from a highly efficient RG flow in an ultraviolet expansion}\label{explicit-map}

In the previous section, we have argued that full diffeomorphism invariance of the dual classical gravity equations is a \textit{necessary} condition for them be rewritten into a highly efficient RG flow. Particularly the effective metric $g_{\mu\nu}(\Lambda)$ in (\ref{schematic2}), which is a functional of the effective single trace operator $t^\mu_{\phantom{\mu}\nu}(\Lambda)$ in a fixed background, must lead to a bulk metric (\ref{FGmetric}) that satisfies the equations of a full-diffeomorphism invariant pure gravity theory. In this section, we will show that diffeomorphism invariance of the classical gravity theory is also a \textit{sufficient} condition for it to be mapped to a highly efficient RG flow. 
In subsection \ref{geometric-variables}, we will show how one can map variables in gravity to those of a highly efficient RG flow. In subsection \ref{calculating-map}, we will show how we can construct the highly efficient RG flow equations explicitly and recover the dual gravity theory from them.

Although our results will be valid for any diffeomorphism invariant classical gravity theory, the explicit calculations presented here will be for the case of Einstein's gravity.

\subsection{Geometric variables and the RG flow}\label{geometric-variables}

The RG flow in a field theory in $d-$dimensions can be geometrically represented very efficiently via a hypersurface foliation $\Sigma_r$ and a congruence of curves $\mathcal{C}_x$ in a $(d+1)-$dimensional spacetime. 

The boundary of this higher dimensional spacetime represents the UV fixed point of the field theory. Let us foliate this spacetime by non-intersecting timelike hypersurfaces $\Sigma_r$, which are labeled uniquely by the radial coordinate $r$ (i.e. there exists a function $r$ such that each hypersurface in the foliation is given by $r=constant$, and the normal to each hypersurface is spacelike). 

Let us also consider a congruence of curves $\mathcal{C}_x$, such that each curve in $\mathcal{C}_x$ intersects each hypersurface in $\Sigma_r$ only once, and furthermore these curves do not intersect with each other. In such a case, we can label each curve in $\mathcal{C}_x$ uniquely by a $d$-dimensional spacetime coordinate $x^\mu$. In a QFT, the local operators are also labeled by the $d-$dimensional spacetime coordinates $x^\mu$ -- so we will call these the field-theory coordinates.

The radial coordinate $r$ can be related to the scale $\Lambda$ by requiring that it is independent of (i) the field theory state, (ii) $l$, the asymptotic curvature radius of the spacetime, and (iii) the field-theory coordinates $x^\mu$.  Then simple dimensional analysis implies that $r=\Lambda^{-1}$. The boundary of the spacetime then corresponds to $r=0$, where the UV data should be specified.  We will repeatedly use the feature of \textit{state-independence} of the RG flow equation (\ref{schematic1}) while mapping the field-theoretic variables to quantities on the gravity side. We will also repeatedly use the fact that $l$, the asymptotic curvature radius, does not appear explicitly in the RG flow equation (\ref{schematic1}), as is evident when it is written in the form (\ref{t-evol-exp}).

The RG flow of the QFT local operators then occurs geometrically along $\mathcal{C}_x$, each point in the curve representing a specific scale $\Lambda$, corresponding to the hypersurface in the foliation $\Sigma_r$ that intersects at that point. 

\paragraph{ADM variables:}If the RG flow maps to a classical theory of gravity, then the description of the RG flow via $(\Sigma_r, \mathcal{C}_x)$ gains a real significance. As stated above, there exists a natural $(d+1)-$dimensional coordinate system associated with $(\Sigma_r, \mathcal{C}_x)$ which is given by the radial coordinate $r$ and the field-theory coordinates $x^\mu$. In this coordinate system, the $(d+1)-$ dimensional metric assumes the general ADM form \cite{Arnowitt:1962hi}:
\begin{equation}\label{ADM}
{\rm d}s^2 = \alpha^2(r,x) {\rm d}r^2 + \gamma_{\mu\nu}(r,x)\left({\rm d}x^\mu + \beta^\mu(x,r) {\rm d}r\right)\left({\rm d}x^\nu + \beta^\nu(r,x) {\rm d}r\right).
\end{equation}
It can be shown that $\gamma_{\mu\nu}(r,x)$ is the induced metric on the hypersurface $\Sigma_r$ and constitutes the dynamical variable in general relativity. The other variables $\alpha(r,x)$ and $\beta^\mu(r,x)$ are non-dynamical, since their radial derivatives do not appear in the equations of gravity, and we call them the pseudo-lapse function and the pseudo-shift vector, respectively (the qualifier \textit{pseudo} is used to distinguish them from the standard case when they are discussed in the context of time-evolution as opposed to radial evolution). They play a role similar to Lagrange multipliers in enforcing a vector-like momentum constraint and a scalar-like Hamiltonian constraint for the data on each hypersurface $\Sigma_r$. Indeed, a choice of $(d+1)-$dimensional coordinate system is related to a specific method of gauge fixing of the diffeomorphism symmetry, and hence specific conditions that determine $\alpha(r,x)$ and $\beta^\mu(r,x)$.
%Strictly speaking, these statements are true for a tubelike region in the bulk spacetime, ending on a sufficiently small patch on the boundary $r=0$. It should be interesting to investigate, how the absence of any global spacelike hypersurface foliation reflects structurally in the RG flow equation, as for instance eq. \ref{t-rg-example}. 

Furthermore, as is well known, in order to describe the RG flow in a CFT, we need asymptotically anti-de Sitter ($AdS$) spacetimes corresponding to the states in the CFT.  The asymptotic isometries of the spacetime map to the conformal group $SO(d, 2)$ of the $d-$dimensional CFT, which is necessary for reproducing the conformal Ward identities of the CFT via equations of gravity \cite{Henningson:1998gx,Henningson:1998ey,Balasubramanian:1999re}. 

An example of the choice of gauge-fixing of bulk diffeomorphisms is the Fefferman-Graham gauge, in which any $(d+1)-$dimensional asymptotically $AdS$ spacetime metric assumes the form (\ref{FGmetric}), meaning that 
\begin{equation}\label{FG}
\alpha = \frac{l}{r}, \quad \beta^\mu = 0, \quad\text{and}\quad
\gamma_{\mu\nu} = \frac{l^2}{r^2}\cdot g_{\mu\nu}.
\end{equation}
The constant $l$ is the radius of the asymptotic $AdS$ region of the spacetime. The corresponding coordinates and $(\Sigma_r, \mathcal{C}_x)$ are well defined for sufficiently small $r$ in any asymptotically $AdS$ spacetime, meaning for sufficiently large $\Lambda$ in the corresponding highly efficient RG flow. This is therefore very suitable for studying the ultraviolet expansion of the RG flow, as we will see soon.

The momentum constraints in Einstein's gravity associated with each hypersurface in $\Sigma_r$ imply that there should be an appropriate quasi-local energy-momentum tensor ${T^\mu_{\phantom{\mu}\nu}}^{{\rm ql}}$ \cite{PhysRevD.47.1407} which is a functional of $\alpha$, $\beta^\mu$, $\gamma_{\mu\nu}$ and $\partial\gamma_{\mu\nu}/\partial r$ and is locally conserved, i.e. it satisfies,
\begin{equation}\label{mom-constraint-gravity}
\nabla_{(\gamma)\mu}{T^\mu_{\phantom{\mu}\nu}}^{{\rm ql}} = 0.
\end{equation}
In case of Einstein's equations, this ${T^\mu_{\phantom{\mu}\nu}}^{{\rm ql}}$ is the well-known Brown-York tensor, as given by \cite{PhysRevD.47.1407}
\begin{equation}\label{Brown-York}
{T^\mu_{\phantom{\mu}\nu}}^{{\rm ql}} = -\frac{1}{8 \pi G_N} \, \gamma^{\mu\rho}\left(K_{\rho\nu}- K \gamma_{\rho\nu}\right).
\end{equation}
Here, $G_N$ is the $(d+1)-$dimensional Newton's gravitational constant, while $K_{\mu\nu}$ is the extrinsic curvature of the hypersurface defined via
\begin{equation}\label{extrinsic-curvature}
K_{\mu\nu} = -\frac{1}{2\alpha}\left(\frac{\partial \gamma_{\mu\nu}}{\partial r} - \nabla_{(\gamma)\mu} \beta_\nu -\nabla_{(\gamma)\nu} \beta_\mu \right),
\end{equation}
with $\beta_\rho = \gamma_{\rho\mu} \beta^\mu$, and $K = K_{\mu\nu}\gamma^{\mu\nu}$. As $\alpha$ and $\beta^\mu$ are non-dynamical, and since they are determined by the gauge-fixing conditions (associated with the coordinate system $r,x^\mu$ and corresponding $(\Sigma_r, \mathcal{C}_x)$), ${T^\mu_{\phantom{\mu}\nu}}^{\rm ql}$ should be regarded as a functional of the induced metric $\gamma_{\mu\nu}$. 

The existence of a quasi-local stress tensor ${T^\mu_{\phantom{\mu}\nu}}^{{\rm ql}}$ that satisfies (\ref{mom-constraint-gravity}) is not an exclusive property of Einstein's gravity. Indeed, it can be shown that such a quasi-local stress tensor exists in any \textit{pure} classical gravity theory that satisfies $(d+1)-$diffeomorphism invariance \cite{Deruelle:2007pt} (see also \cite{Balcerzak:2007da}) by virtue of the fact that the classical gravity equations satisfy $(d+1)-$Bianchi identities. Furthermore, this quasi-local stress tensor should be unique \textit{on-shell} up to terms which are conserved identically (for explicit constructions in some higher derivative gravity theories see \cite{Balcerzak:2007da, Deruelle:2007pt}). Indeed, if there is one more such non-trivial quasi-local stress tensor which is conserved in an arbitrary solution and does not differ from the standard one by terms which are trivially conserved by $d-$dimensional Bianchi identities, it will automatically imply the existence of more constraints which are compatible with the radial evolution in addition to those allowed by $(d+1)-$diffeomorphism invariance. This will be inconsistent with the count of allowed propagating degrees of freedom in the generic case. 

It is intuitively obvious that in order to map classical gravity equations to a highly efficient RG flow, we need to identify $\gamma_{\mu\nu}$ with $g_{\mu\nu}$ and ${T^\mu_{\phantom{\mu}\nu}}^{\rm ql}$ with $t^\mu_{\phantom{\mu}\nu}$ at $r=\Lambda^{-1}$. This turns out to be a bit naive, because $\gamma_{\mu\nu}$ and ${T^\mu_{\phantom{\mu}\nu}}^{\rm ql}$ both blow up in the UV, meaning at $r=0$. This blow up indeed has an immense physical significance, allowing us to map Weyl transformations of the field theory data to specific diffeomorphisms in the corresponding $(d+1)-$dimensional spacetime geometry. We will discuss this aspect in more detail in Section \ref{lifted-Weyl} and Section \ref{decipher}.

\paragraph{The (dis)appearance of the parameter $l$:} The parameter $l$, which is the asymptotic curvature radius of the $(d+1)-$dimensional spacetime, by itself has no meaning in the CFT. It is basically the unit of measurement for mass/length/time on the gravitational side -- dimensionless quantities in these units correspond to parameters in the CFT. The asymptotic curvature radius $l$ is determined by the (negative) cosmological constant in the  $(d+1)-$dimensional gravity.

The other parameters of the classical gravity theory are of two kinds:
\begin{itemize}
\item the \textit{overall factor} $1/(16\pi G_N)$ which multiplies the $(d+1)-$dimensional gravity action, and
\item the \textit{relative} coefficients giving higher derivative corrections to Einstein's gravity, involving the $(d+1)-$dimensional Riemann tensor and its covariant derivatives -- without loss of generality it can be assumed that these relative coefficients should take the form of numerical constant times ${\alpha^\prime_{(i)}}^n$, for a suitable $n$ and with  $\alpha^\prime_{(i)}$ being a parameter with mass dimension $-2$ (if the theory of gravity is a low energy limit of a string theory, then there is only one $\alpha'$, i.e. the inverse of the string tension).
\end{itemize}

In the large $N$ limit, where the factorisation of  expectation values of multi-trace operator applies, the highly efficient RG flow equation, like (\ref{t-rg-example})  can be thought of as a classical equation that can be mapped to classical gravity as discussed in the Introduction. Thus the factor $N^2$ should be related to a parameter that controls quantum corrections on the gravity side, which is $1/(16\pi G_N)$ -- the corresponding dimensionless parameter is $l^{d-1}/(16\pi G_N)$. Therefore, we expect that
\begin{equation}
N^2 \approx \frac{l^{d-1}}{16\pi G_N}.
\end{equation}
In the case of $\mathcal{N} = 4$ SYM holography, $N$ is the rank of the gauge group and $l^3/(16\pi G_5) = N^2/(8\pi^2)$ \cite{Maldacena:1997re}.

Furthermore, we expect that only in the strongly interacting limit, where the algebra of operators is possibly generated by a \textit{finite} number of single-trace operators, we can construct effective physics via a \textit{first order} RG flow involving a \textit{finite} number of single-trace operators. Going away from the strongly interacting limit implies that we need to either include infinitely many single-trace operators, or include higher derivative effects in the RG flow when we integrate out single-trace operators, whose anomalous dimensions are parametrically large as functions of the couplings,  and study the scale-evolution of the surviving single trace operators. Nevertheless, if we do systematic perturbation theory (in inverse powers of the couplings) about the infinite strong coupling limit, we can retain the feature that the RG flow of the finite number of single-trace operators with small anomalous dimensions is still first order -- the higher derivative effects can be removed by substituting leading or subleading order solutions as is often done in perturbative analysis. We can expect that this feature should be replicated when we map a highly efficient RG flow to a classical gravity theory.

In the sector of states where we consistently have only $t^\mu_{\phantom{\mu}\nu}$ as the evolving single-trace operator, the dual description is given by pure gravity. The dual theory of gravity however describes the evolution of the effective metric $g_{\mu\nu}(\Lambda)$, which in addition to $t^\mu_{\phantom{\mu}\nu}(\Lambda)$ requires another piece of information (or rather integration constant) to be specified uniquely, namely the fixed background metric on which the field theory lives. Therefore its evolution should be given by a two-derivative equation. Indeed the unique diffeomorphism invariant two-derivative pure gravity theory is Einstein's gravity with a cosmological constant. Departure from the infinite strong coupling limit, should induce higher derivative corrections both in the RG flow and the dual gravity theory. Nevertheless, if we want to preserve the scale-evolution of the RG flow as first order, we need to also treat the higher derivative corrections on the gravity side perturbatively too when we map the RG flow to a dual gravity theory.

We thus expect that the dimensionless parameters $\alpha'_{(i)}/l^2$ giving corrections to Einstein's gravity are related to functions of the (dimensionless) coupling constants $\lambda_{(i)}$ in the CFT, whose values are small in the strongly interacting limit. In other words, we expect that
\begin{equation}
\frac{\alpha'_{(i)}}{l^2} \approx f\left(\lambda_{(i)}\right), \quad \text{such that $f\rightarrow 0$ as $\lambda_{(i)}\rightarrow\infty$.}
\end{equation}
We have assumed here that the CFT has a weakly coupled quasi-particle like description when all $\lambda_{(i)}$ are small. In $\mathcal{N}=4$ SYM, there is only one coupling constant, which is the 't Hooft coupling $\lambda$. This is related holographically to the string tension $\alpha'$ that gives higher derivative corrections to Einstein's gravity via \cite{Maldacena:1997re}
\begin{equation}
\frac{\alpha'}{l^2} = \sqrt{\frac{1}{\lambda}}.
\end{equation} 

We want to establish a relation between $\gamma_{\mu\nu}$ and $g_{\mu\nu}$, and between ${T^\mu_{\phantom{\mu}\nu}}^{\rm ql}$ and $t^\mu_{\phantom{\mu}\nu}$, such that $l$ does not appear explicitly in the RG flow equation, but only in the dimensionless combinations $l^{d-1}/(16 \pi G_N)$ and $\alpha'_{(i)}/{l^2}$, which as discussed above are related directly to parameters in the CFT. 

\paragraph{Relationship between $\gamma_{\mu\nu}$ and $g_{\mu\nu}$:} Let us begin with establishing the relation between $\gamma_{\mu\nu}$ and $g_{\mu\nu}$. In order to preserve the effective form of the Ward identity (\ref{WILambda}), it is natural to impose 
\begin{equation}\label{gamma-match}
\nabla_{(\gamma)} = \nabla_{(g)},
\end{equation}
which then implies
\begin{equation}\label{gandgammahyp}
g_{\mu\nu} = f\left(\frac{r}{l}\right)\cdot \gamma_{\mu\nu}
\end{equation}
at $r=\Lambda^{-1}$, \textit{if the relationship between $\gamma_{\mu\nu}$ and $g_{\mu\nu}$ is to be state-independent}, which also demands that (\ref{gamma-match}) should be valid in an arbitrary gravity solution as well. The latter also follows from (\ref{gandgammahyp}).

The factor $f(r/l)$ can in principle depend on the auxiliary $\alpha$ and $\beta^\mu$, but not on the state-dependent variables which appear in the dual gravity solutions. However, depending on the choice of gauge fixing, $\alpha$ and $\beta^\mu$, may be functions of both $r$ and $x^\mu$. Therefore, such a dependence should not exist.

Furthermore, requiring that the asymptotic curvature radius $l$ does not appear explicitly in the scale evolution of $g_{\mu\nu}$, and that at $\Lambda =\infty$, i.e. $r=0$, $g_{\mu\nu}$ coincides with the background metric on which the field theory lives, we readily obtain that
\begin{equation}\label{gandgamma}
g_{\mu\nu} = \frac{r^2}{l^2}\cdot\gamma_{\mu\nu}
\end{equation}
at $r=\Lambda^{-1}$. To see that the scale evolution of $g_{\mu\nu}$ as obtained via the above relation does not depend on $l$ explicitly, we need to study the dynamics, which will be done in the next subsection. 

The relation \eqref{gandgamma} also implies the traditional rule of holographic correspondence, namely that
\begin{equation}
g^{\rm (b)}_{\mu\nu}  = \lim_{r\rightarrow0}\,\,\frac{r^2}{l^2}\cdot\gamma_{\mu\nu} 
\end{equation}
coincides with the background metric on which the CFT lives. In the literature, $g^{\rm (b)}_{\mu\nu}$ is often called the \textit{boundary metric} of the $(d+1)-$dimensional spacetime.

In the Fefferman-Graham gauge, the relation \eqref{gandgamma} reduces to the identification of 
$g_{\mu\nu}(\Lambda)$ with that which appears in the bulk metric (\ref{FGmetric}).

Notably, the $(d+1)-$dimensional spacetime corresponding to the vacuum of the CFT in flat Minkowski space is pure $AdS_{d+1}$. In the Fefferman-Graham gauge, $g_{\mu\nu}= \eta_{\mu\nu}$ for all values of $r$ in pure $AdS_{d+1}$.  This implies that $g_{\mu\nu}$ remains $\eta_{\mu\nu}$ at all scales in the RG flow. This is expected -- since $\langle {t^\mu_{\phantom{\mu}\nu}}^\infty\rangle$ vanishes in the vacuum, so should  $\langle t^\mu_{\phantom{\mu}\nu}(\Lambda)\rangle$ and $g_{\mu\nu}(\Lambda)$.

\paragraph{Relationship between} ${T^\mu_{\phantom{\mu}\nu}}^{\rm ql}$ \text{and} $t^\mu_{\phantom{\mu}\nu}$: The next step is to establish the relation between ${T^\mu_{\phantom{\mu}\nu}}^{\rm ql}$ on the gravity side with $t^\mu_{\phantom{\mu}\nu}$ of the highly efficient RG flow at $r=\Lambda^{-1}$. Once again, in order to preserve the effective Ward identity (\ref{WILambda}) at all scales, it is natural to postulate that the relationship between ${T^\mu_{\phantom{\mu}\nu}}^{\rm ql}$ and $t^\mu_{\phantom{\mu}\nu}$ should be of the form
\begin{equation}\label{tandT}
t^\mu_{\phantom{\mu}\nu} =h\left( \frac{r}{l}\right)\cdot\left({T^\mu_{\phantom{\mu}\nu}}^{\rm ql} +{T^\mu_{\phantom{\mu}\nu}}^{\rm ct}\right),
\end{equation}
where ${T^\mu_{\phantom{\mu}\nu}}^{\rm ct}$ denotes general counterterms which are \textit{conserved identically}, meaning that they satisfy $\nabla_{(\gamma)\mu}{T^\mu_{\phantom{\mu}\nu}}^{\rm ct} =0$ when considered individually, via $d-$dimensional Bianchi identities. The latter thus assumes the general form:
\begin{eqnarray}\label{Tct}
{T^\mu_{\phantom{\mu}\nu}}^{\rm ct} &=&-\frac{1}{8\pi G_N} \Bigg[C_{(0)}\left( \frac{r}{l}, \frac{\alpha'_{(i)}}{l^2}\right)\cdot \frac{1}{l} \cdot \delta^\mu_{\phantom{\mu}\nu} + C_{(2)}\left( \frac{r}{l}, \frac{\alpha'_{(i)}}{l^2}\right)\cdot l \cdot \left(R^\mu_{\phantom{\mu}\nu}[\gamma] - \frac{1}{2}R[\gamma]\delta^\mu_{\phantom{\mu}\nu}\right) +
 \cdots\,\Bigg].
\end{eqnarray}
We note that each term in ${T^\mu_{\phantom{\mu}\nu}}^{\rm ct}$ above, namely the one proportional to $\delta^\mu_{\phantom{\mu}\nu}$ and the one proportional to the Einstein tensor constructed out of $\gamma_{\mu\nu}$, are the unique terms up to two derivatives, which are conserved identically. In order to be conserved identically (in other words to be conserved even off-shell), we require the terms in ${T^\mu_{\phantom{\mu}\nu}}^{\rm ct}$ to be functionals of $\gamma_{\mu\nu}$ only, up to proportionality factors that are functions of $r$ only (i.e. independent of $x^\mu$). Therefore, these terms cannot depend on the auxiliary ADM variables, namely $\alpha$ and $\beta^\mu$. Up to any given order in derivatives, there are only a finite number of such terms, and there is a well-known procedure to construct them \footnote{This procedure is to take all possible diffeomorphism invariant Lagrangian densities (like $R$) up to fixed order in derivatives, and then construct the corresponding equations of motion (like the Einstein tensor for $R$). The diffeomorphism invariance of the Lagrangian implies the Bianchi identities.}.

This argument for construction of counterterms does not rely on any choice of boundary condition in gravity at the cut-off scale (hypersurface) as in the case of traditional holographic renormalisation \cite{Akhmedov:1998vf,Henningson:1998gx,Henningson:1998ey,Balasubramanian:1999re,deBoer:1999xf,deHaro:2000xn} (which assumes the Dirichlet boundary condition at any cut-off scale). The basic feature that only terms which are conserved identically can appear in the counterterms is simply a consequence of the fact that ${T^\mu_{\phantom{\mu}\nu}}^{\rm ql}$ is the only non-trivial tensor which is conserved via classical gravity equations of motion in all solutions of gravity. However, ${T^\mu_{\phantom{\mu}\nu}}^{\rm ql}$ is not conserved identically, implying that it is the only non-trivial tensor that is  conserved \textit{only} on-shell in an \textit{arbitrary solution}. Thus assuming that the relationship between ${T^\mu_{\phantom{\mu}\nu}}^{\rm ql}$ and $t^\mu_{\phantom{\mu}\nu}$ is independent of the state in the CFT, no tensor other than ${T^\mu_{\phantom{\mu}\nu}}^{\rm ql}$ which is not conserved identically can appear in \eqref{tandT}. Therefore, the countertems in ${T^\mu_{\phantom{\mu}\nu}}^{\rm ct}$ have to be conserved identically.

%This argument implies that even if we make a specific choice of a specific boundary condition at the cut-off to construct a counterterm tensor ${T^\mu_{\phantom{\mu}\nu}}^{\rm ct}$ which respects the corresponding variational principle, it has to take the form \eqref{Tct} on-shell.

We can determine the $r/l$ dependence in the various coefficient functions $h(r/l)$ or $C_{(i)}(r/l, \alpha'_{(i)}/l^2)$ in \eqref{tandT} and \eqref{Tct} by requiring that $t^\mu_{\phantom{\mu}\nu}$ does not depend on $l$ explicitly. This implies that
\begin{equation}\label{handCi}
h\left(\frac{r}{l}\right) = \left(\frac{l}{r}\right)^d, \quad \text{$C_{(i)}\left(\frac{r}{l}, \frac{\alpha'_{(i)}}{l^2}\right) = C_{(i)}\left(\frac{\alpha'_{(i)}}{l^2}\right)$ and thus independent of $\frac{r}{l}$.} 
\end{equation}
A finite number of these constants $C_{(i)}(\alpha'_{(i)}/l^2)$ can be determined by requiring that $t^\mu_{\phantom{\mu}\nu}(\Lambda=\infty) = \langle{t^\mu_{\phantom{\mu}\nu}}^\infty\rangle$ is finite \cite{Henningson:1998gx,Henningson:1998ey,Balasubramanian:1999re,deBoer:1999xf,deHaro:2000xn}. As discussed before and as will be shown explicitly in the second part of our work \cite{secondpart} based on the results of \cite{Kuperstein:2013hqa}, these counterterms can also be fixed by the infrared criterion that the endpoint of the RG flow should be transformed to a non-relativistic fixed point with a finite number of parameters under the rescaling (\ref{rescaling}).

Since the form of $t^\mu_{\phantom{\mu}\nu}$ as given by \eqref{tandT} and \eqref{Tct} is gauge-independent, we can check whether the explicit form satisfies all the criteria discussed above in Fefferman-Graham gauge for the sake of convenience. In the Section \ref{diffeo-action}, we will then see explicitly that the above features are also preserved under infinitesimal gauge transformations. Let us examine the case of Einstein's equations first. Here, the counterterm coefficients $C_{(i)}$ should be just pure universal numbers (independent of the underlying CFT, which is infinitely strongly interacting).

In the Fefferman-Graham gauge \eqref{FG}, we obtain
\begin{equation}\label{KmunuFG}
K_{\mu\nu} = \frac{l}{r}\left(\frac{g_{\mu\nu}}{r}-\frac{1}{2} \frac{\partial g_{\mu\nu}}{\partial r}\right).
\end{equation}
Let us also define $z^\mu_{\phantom{\mu}\nu}$ as
\begin{equation}\label{z}
z^\mu_{\phantom{\mu}\nu} = g^{\mu\rho}\cdot\frac{\partial g_{\rho\nu}}{\partial r}.
\end{equation}
We observe that \eqref{gandgamma} implies
\begin{equation*}
R^\mu_{\phantom{\mu}\nu\rho\sigma}[\gamma] = R^\mu_{\phantom{\mu}\nu\rho\sigma}[g].
\end{equation*}
Using the above observation, we can show that \eqref{Brown-York}, \eqref{tandT}, \eqref{Tct}, \eqref{handCi} and \eqref{KmunuFG} imply that in the case of Einstein's gravity
\begin{eqnarray}\label{tFG}
t^\mu_{\phantom{\mu}\nu} &=& \frac{l^{d-1}}{16\pi G_N}\Bigg[\frac{1}{r^{d-1}}\cdot \left(z^\mu_{\phantom{\mu}\nu} - ({\rm Tr}\, z) \,\delta^\mu_{\phantom{\mu}\nu}\right) +2\cdot \frac{1}{r^d}\cdot  \left( d- 1 - C_{(0)}\right) \cdot \delta^\mu_{\phantom{\mu}\nu}-\nonumber\\&&
\quad\quad\quad\quad - 2\cdot \frac{1}{r^{d-2}}\cdot C_{(2)}\cdot \left(R^\mu_{\phantom{\mu}\nu}[g] - \frac{1}{2}R[g]\delta^\mu_{\phantom{\mu}\nu}\right)
+ \cdots \Bigg],
\end{eqnarray}
at $r=\Lambda^{-1}$. As mentioned earlier (and as we will also prove in the next subsection), $g_{\mu\nu}$ does not depend on $l$ explicitly (or even implicitly in the case of Einstein's gravity). Then it follows from the above that $t^\mu_{\phantom{\mu}\nu}$ indeed depends on $l$ only through the dimensionless combination $l^{d-1}/(16\pi G_N)$ in Einstein's gravity, as claimed. Furthermore, the terms not shown, are subleading in the ultraviolet expansion in $\Lambda^{-1}$, in which it can be systematically expanded, as we will see in the next subsection. We will demonstrate additionally that \eqref{handCi} remains applicable for higher derivative gravity also, but in this case, $C_{(i)}$ will also depend on the parameters $\alpha'_{(i)}/l^2$.

In order for $t^\mu_{\phantom{\mu}\nu}$ to be finite at $\Lambda=\infty$, we need \cite{Henningson:1998gx,Henningson:1998ey,Balasubramanian:1999re}
\begin{equation}\label{C0C2}
C_{(0)} = d -1, \quad C_{(2)} = - \frac{1}{d-2}, \quad \text{etc.}
\end{equation}
for $d>2$ in Einstein's gravity. These coefficients can be readily determined by substituting the on-shell ultraviolet expansion of $g_{\mu\nu}$ (which is also known as the Fefferman-Graham expansion \cite{FG:1984} in gravity) into \eqref{tFG}, as will be presented in the next subsection. Since $t^\mu_{\phantom{\mu}\nu}$ has only a finite number of $\Lambda^{n}$ (or equivalently $r^{-n}$) UV divergences, only a finite number of counterterm coefficients $C_{(i)}$ can be determined from ultraviolet finiteness.

There is however one ambiguity still remaining -- namely that of the choice of overall normalisation by a numerical constant in (\ref{tandT}) which maps the gravity variables to the $t^\mu_{\phantom{\mu}\nu}$ of the RG flow. This can be readily fixed by studying the two-point correlation function for instance, which can be reproduced by our constructive field theory approach as will be discussed in \cite{secondpart}. We will make a simple convenient choice of the normalisation soon.

Once this normalisation is fixed, the crucial result is that \textit{our infrared criterion for the end point of the RG flow establishes a unique map between variables of a $(d+1)-$diffeomorphism invariant pure gravity theory and the $t^\mu_{\phantom{\mu}\nu}(\Lambda)$ in a $d-$dimensional highly efficient RG flow}. This assertion will be fully established with explicit examples in \cite{secondpart}. Furthermore, we will also establish that the infrared criterion also determines the UV data, which is otherwise obtained via absence of naked singularities in the solutions of gravity in the traditional holographic correspondence.

It is indeed gratifying that we need not study the UV behaviour in order to establish this map between gravity and the RG flow, or to determine the UV data. Since the RG flow is first order, the infrared criterion should suffice to see how the theory of gravity can be mapped to the effective dual field theory observables. In particular, this implies that our procedure can be implemented even when the theory of gravity does not lead to well defined asymptotic behaviour as in asymptotically $AdS$ spacetimes. In this case, we need to implement a UV cut-off and still the map between gravity and the RG flow, and the UV data at the cut-off can be fully determined in a unique way by the infrared criterion. This observation will be utilised in \cite{secondpart} to propose new non-perturbative frameworks for theories like QCD, via a proposal for matching perturbative Wilsonian RG flow in the UV with our highly efficient RG flow reconstruction of IR physics, as concrete applications of our approach for reformulation of holography.

\paragraph{Effect of the conformal anomaly:} It is convenient to rescale $t^\mu_{\phantom{\mu}\nu}$ by the factor $(16\pi G_N)/l^{d-1}$, i.e. to perform the replacement
\begin{equation}\label{rescaling}
t^\mu_{\phantom{\mu}\nu} \rightarrow \frac{16\pi G_N}{l^{d-1}}t^\mu_{\phantom{\mu}\nu}.
\end{equation}
This amounts to rescaling by a factor of $\propto 1/N^2$, and this rescaled $t^\mu_{\phantom{\mu}\nu}$ is finite in the large $N$ limit. In $\mathcal{N}=4$ SYM holography, this rescaling factor is precisely $8\pi^2/N^2$.

The conformal anomaly does not contribute to ${\rm Tr}\, t^\infty$ in flat Minkowski space background, but not surprisingly it has an effect on the RG flow. In order to see this, let us assume minimalistic counter-terms required to cancel UV divergences only. Of course, this may not necessarily be justified, since other non-minimalistic \textit{marginal} counter-terms (proportional to $R^2$ in $d=4$ and multiplied by a constant instead of $\log r$) may be required to satisfy our infrared criterion. However, one way to see whether our assumption is indeed valid will be to go to fourth order in the hydrodynamic derivative expansion and repeat the calculations in  \cite{Kuperstein:2013hqa} to see for which values of the non-minimalistic marginal counterterm coefficients the end point of the RG flow can be transformed to the incompressible non-relativistic Navier-Stokes fixed point under the rescaling (\ref{rescaling}). This is beyond the scope of this paper. Therefore we will simply proceed with the assumption that the non-minimalist marginal counterterms vanish -- of course explicit calculations in the future can easily correct the resulting equations (this ambiguity will not affect the results in \cite{secondpart}, the second part of our work because we will restrict ourselves to second order in the hydrodynamic derivative expansion).

In this case, in $d=4$ and when the RG flow maps to Einstein's gravity, we can write \cite{deHaro:2000xn} \footnote{In comparing our equation with this reference, we must keep in mind that the latter uses the convention for definition of the Riemann tensor in which $AdS$ has constant positive curvature.}
\begin{eqnarray}
t^\mu_{\phantom{\mu}\nu} &=& {t^\mu_{\phantom{\mu}\nu}}^{\rm bare} + {t^\mu_{\phantom{\mu}\nu}}^{\rm ct(1)} + {t^\mu_{\phantom{\mu}\nu}}^{\rm ct(2)} + {t^\mu_{\phantom{\mu}\nu}}^{\rm ct(a)}\nonumber\\&& + \text{terms that will contribute to RG flow eq. at $\mathcal{O}(\Lambda^{-7} \log \Lambda)$},
\end{eqnarray}
where
\begin{eqnarray}
{t^\mu_{\phantom{\mu}\nu}}^{\rm bare} &=& -2\frac{l}{r^4} \, \gamma^{\mu\rho}\left(K_{\rho\nu}- K \gamma_{\rho\nu}\right),\\
{t^\mu_{\phantom{\mu}\nu}}^{\rm ct(1)} &=&  - 6 \frac{1}{r^4}\delta^\mu_{\phantom{\mu}\nu},\\
{t^\mu_{\phantom{\mu}\nu}}^{\rm ct(2)} &=&   \frac{l^2}{r^4}\left(R^\mu_{\phantom{\mu}\nu}[\gamma] - \frac{1}{2}R[\gamma]\delta^\mu_{\phantom{\mu}\nu}\right),\\
{t^\mu_{\phantom{\mu}\nu}}^{\rm ct(a)} &=&- \frac{l^4}{r^4}\log \, r \Bigg(\frac{1}{8}R^\mu_{\phantom{\mu}\alpha\nu\beta}[\gamma]R^{\alpha\beta}[\gamma]- \frac{1}{48}\nabla^\mu\nabla_\nu R[\gamma]+ \frac{1}{16}\nabla^2 R^{\mu}_{\phantom{\mu}\nu}[\gamma]-\frac{1}{24}R[\gamma]R^{\mu}_{\phantom{\mu}\nu}[\gamma]+\nonumber\\&&\qquad\qquad+\left(\frac{1}{96}R^2[\gamma]-\frac{1}{32}R_{\alpha\beta}[\gamma]R^{\alpha\beta}[\gamma]-\frac{1}{96}\nabla^2 R[\gamma]\right)\delta^{\mu}_{\phantom{\mu}\nu}\Bigg).
\end{eqnarray}
In Fefferman-Graham gauge the above reads
\begin{eqnarray}\label{t-z4d}
t^\mu_{\phantom{\mu}\nu} &=& \frac{1}{r^3}\cdot \left(z^\mu_{\phantom{\mu}\nu} - ({\rm Tr}\, z) \,\delta^\mu_{\phantom{\mu}\nu}\right) + \frac{1}{r^2} \left(R^\mu_{\phantom{\mu}\nu}[g] - \frac{1}{2}R[g]\delta^\mu_{\phantom{\mu}\nu}\right)
+ \nonumber\\&& 
- \log r\Bigg(\frac{1}{8}R^\mu_{\phantom{\mu}\alpha\nu\beta}[g]R^{\alpha\beta}[g]- \frac{1}{48}\nabla^\mu\nabla_\nu R[g]+ \frac{1}{16}\nabla^2 R^{\mu}_{\phantom{\mu}\nu}[g]-\frac{1}{24}R[g]R^{\mu}_{\phantom{\mu}\nu}[g]+\nonumber\\&&\qquad\qquad+\left(\frac{1}{96}R^2[g]-\frac{1}{32}R_{\alpha\beta}[g]R^{\alpha\beta}[g]-\frac{1}{96}\nabla^2 R[g]\right)\delta^{\mu}_{\phantom{\mu}\nu}\Bigg) + \cdots .
\end{eqnarray}
The $\log$ counterterm ${t^\mu_{\phantom{\mu}\nu}}^{\rm ct(a)}$ is completely fixed by the conformal anomaly, in particular the central charges.

As we will see soon, only one possible non-minimalist counter-term, namely $\nabla^2 R^\mu_{\phantom{\mu}\nu}$ without the $\log r$ pre-factor, can affect the highly efficient RG flow equation \eqref{t-rg-example} up to $\mathcal{O}(1/\Lambda^7 \, \log \, \Lambda)$.

\subsection{The highly efficient RG flow from gravity equations and vice versa}\label{calculating-map}

\paragraph{Mapping in the UV expansion:} Assuming that $g_{\mu\nu}(\Lambda)$ depends only  on $t^\mu_{\phantom{\mu}\nu}(\Lambda)$ and $\Lambda$ explicitly one can do an UV expansion, which is valid when $\langle t^\mu_{\phantom{\mu}\nu}(\Lambda)\rangle/ \Lambda^d \ll 1$. These conditions should be valid for any state in the CFT for sufficiently large $\Lambda$, meaning for $\Lambda \gg \Lambda_{\rm IR}$, with $\Lambda_{\rm IR}$ being a suitable (state-dependent) infrared scale. 

It is actually convenient to first write this expansion in terms of ${t^\mu_{\phantom{\mu}\nu}}^\infty$, because then it matches with the well-known asymptotic Fefferman-Graham expansion \cite{FG:1984} of $g_{\mu\nu}$ in the $(d+1)-$dimensional metric \eqref{FGmetric}. Since in the case of a flat Minkowksi background metric ${t^\mu_{\phantom{\mu}\nu}}^\infty$ satisfies $\partial_\mu {t^\mu_{\phantom{\mu}\nu}}^\infty=0$, and since moreover ${\rm Tr}\, t^\infty = 0$ in a CFT, the most general expansion takes the form:
\begin{eqnarray}\label{g-FG-expansion}
g_{\mu\nu} &=& \eta_{\mu\nu} +c_4\left(\frac{\alpha_{(i)}}{l^2}\right) \cdot\frac{1}{\Lambda^4} \cdot \eta_{\mu\alpha}{t^\alpha_{\phantom{\alpha}\nu}}^\infty
+c_{6}\left(\frac{\alpha_{(i)}}{l^2}\right)\cdot \frac{1}{\Lambda^6}\cdot \eta_{\mu\alpha}\Box {t^\alpha_{\phantom{\alpha}\nu}}^\infty +
\nonumber\\&&
+  \frac{1}{\Lambda^8}\cdot \Big(c_{8a}\left(\frac{\alpha_{(i)}}{l^2}\right)\cdot \eta_{\mu\alpha} {t^\alpha_{\phantom{\alpha}\rho}}^\infty
{t^\rho_{\phantom{\rho}\nu}}^\infty +c_{8b}\left(\frac{\alpha_{(i)}}{l^2}\right)\cdot \eta_{\mu\nu}\cdot {t^\alpha_{\phantom{\alpha}\beta}}^\infty
{t^\beta_{\phantom{\beta}\alpha}}^\infty+ \nonumber\\ &&\quad \quad\quad+c_{8c}\left(\frac{\alpha_{(i)}}{l^2}\right)\cdot\eta_{\mu\alpha}\Box^2 {t^\alpha_{\phantom{\alpha}\nu}}^\infty\Big)+\nonumber\\&&
+ \mathcal{O}\left(\frac{1}{\Lambda^{10}}\right),
\end{eqnarray}
in $d=4$. We have assumed that the conformal anomaly does not introduce any $\log$ term, which will be vindicated by mapping the RG flow to the classical gravity equation. A similar expansion in $d=4$ for $t^\mu_{\phantom{\mu}\nu}(\Lambda)$ reads
\begin{eqnarray}\label{t-FG-expansion}
t^\mu_{\phantom{\mu}\nu} &=& {t^\mu_{\phantom{\mu}\nu}}^\infty
+b_{2}\left(\frac{\alpha_{(i)}}{l^2}\right)\cdot \frac{1}{\Lambda^2}\cdot \Box {t^\mu_{\phantom{\mu}\nu}}^\infty +
\nonumber\\&&
+  \frac{1}{\Lambda^4}\cdot \Big(b_{4a}\left(\frac{\alpha_{(i)}}{l^2}\right)\cdot  {t^\mu_{\phantom{\mu}\rho}}^\infty
{t^\rho_{\phantom{\rho}\nu}}^\infty +b_{4b}\left(\frac{\alpha_{(i)}}{l^2}\right)\cdot \delta^\mu_{\phantom{\mu}\nu} \cdot {t^\alpha_{\phantom{\alpha}\beta}}^\infty
{t^\beta_{\phantom{\beta}\alpha}}^\infty+ \nonumber\\ &&\quad \quad\quad+b_{4c}\left(\frac{\alpha_{(i)}}{l^2}\right)\cdot \Box^2 {t^\mu_{\phantom{\mu}\nu}}^\infty\Big)+\nonumber\\&&
+ \frac{1}{\Lambda^4}\, \log \, \Lambda \cdot \tilde{b}_{4c}\left(\frac{\alpha_{(i)}}{l^2}\right)\cdot \Box^2 {t^\mu_{\phantom{\mu}\nu}}^\infty+ \mathcal{O}\left(\frac{1}{\Lambda^{6}}\log\Lambda\right).
\end{eqnarray}
Above we have assumed that the conformal anomaly restricts the coefficient of $(1/\Lambda^4)\, \log\, \Lambda$ term to be proportional to $\Box^2 {t^\mu_{\phantom{\mu}\nu}}^\infty$ only, disregarding other possible contributions which have the same dimensions. This assumption will be vindicated once again by mapping to classical gravity. 

The coefficients of these expansions are functions of $\alpha'_{(i)}/l^2$, which as noted in the previous subsection are in turn functions of the coupling constants $\lambda_{(i)}$ of the CFT. In the infinitely strongly interacting limit, when the gravitational dynamics is given by Einstein's equation with a negative cosmological constant, these coefficients are pure universal numbers.

Let us first study the case when the highly efficient RG flow maps to Einstein's equation with a negative cosmological constant. It is convenient to start from gravity equations first and then derive the corresponding highly efficient RG flow, and then perform the inverse mapping.

As mentioned before, the asymptotic radius $l$ is related to the $(d+1)-$dimensional cosmological constant $\Lambda_{\rm cc}$, and in Einstein's gravity
\begin{equation}
\Lambda_{\rm cc} = -\frac{d(d-1)}{2l^2}.
\end{equation}
With this convention $(d+1)-$dimensional Einstein's equation for $g_{\mu\nu}$ in the Fefferman-Graham metric \eqref{FGmetric} reduces to (see \cite{deHaro:2000xn} for example):
\begin{eqnarray}
\frac{\partial}{\partial r}{z^\mu_{\phantom{\mu}\nu}} - \frac{d-1}{r} z^\mu_{\phantom{\mu}\nu} + {\rm Tr}\, z\left(\frac{1}{2}z^\mu_{\phantom{\mu}\nu}-\frac{1}{r}\delta^\mu_{\phantom{\mu}\nu}\right)
&=& 2 \,\, R^\mu_{\phantom{\mu}\nu}, \label{tensor-equation} \\
\nabla_\mu\left(z^\mu_{\phantom{\mu}\nu} - ({\rm Tr}\, z) \delta^\mu_{\phantom{\mu}\nu}\right) &=& 0, \label{vector-equation} \\
\frac{\partial}{\partial r}{{\rm Tr}\, z}- \frac{1}{r}{\rm Tr}\, z +\frac{1}{2}{\rm Tr}\, z^2 &=& 0, \label{scalar-equation}
\end{eqnarray}
where $z^\mu_{\phantom{\mu}\nu}$ is as defined in \eqref{z}. Crucially as claimed before, the asymptotic curvature radius $l$ does not appear in these equations which determine the radial evolution of $g_{\mu\nu}$. Thus $g_{\mu\nu}$ does not depend on $l$ as claimed in the previous subsection.

%We are using conventions where the Riemann curvature tensor $R^A_{\phantom{A}BCD}$ is defined via $R^A_{\phantom{A}BCD} = \partial_C \Gamma^A_{BD} + \Gamma^A_{CE}\Gamma^E_{BD} - (C\leftrightarrow D) $. This differs from the convention of \cite{deHaro:2000xn} by a minus sign.
%We are back to old notations where $\nabla$ is the covariant derivative constructed from $g_{\mu\nu}$ and $R^\mu_{\phantom{\mu}\nu}$ is the Ricci tensor also obtained from  $g_{\mu\nu}$.

The vector equation \eqref{vector-equation} is the momentum constraint, and it is easy to see from \eqref{tFG} that it implies the conservation of $t^\mu_{\phantom{\mu}\nu}$. The scalar equation \eqref{scalar-equation} asymptotically implies ${\rm Tr}\, t^\infty =0$ when the boundary metric is flat Minkowski space. Of course, both these equations are constraints -- once they are satisfied at any hypersurface $r = \text{constant}$, the dynamical radial evolution given by \eqref{tensor-equation} preserves them for all values of $r$. 

Substituting the Anstaz \eqref{g-FG-expansion} in \eqref{tensor-equation}, and replacing $\Lambda$ by $r^{-1}$, we can determine all the coefficients in the UV expansion. We obtain \cite{Gupta:2008th}
\begin{equation}\label{cns}
c_4 = \frac{1}{4}, \quad c_6 = - \frac{1}{48}, \quad  c_{8a} = \frac{1}{32}, \quad c_{8b} = -\frac{1}{384}, \quad c_{8c} = \frac{1}{1536}, \quad{\rm etc.}
\end{equation}
in $d=4$. Conveniently all coefficients except $c_{8b}$ can be determined in $d=4$ from the tensor equation \eqref{tensor-equation} alone.  To determine $c_{8b}$, we need to use the scalar constraint \eqref{scalar-equation} too. Indeed, from this point of view an appropriate linear combination of the tensor equation and the scalar equation times $\delta^\mu_{\phantom{\mu}\nu}$ can be regarded as the radial dynamical equation \cite{Gupta:2008th}, which can determine all the coefficients in the UV expansion \eqref{g-FG-expansion}, including $c_{8b}$.

%In order to compare with \cite{Gupta:2008th}, we need to further rescale $t^\mu_{\phantom{\mu}\nu}$ here by a factor of $4$. In this reference, the rescaling factor as in \eqref{rescaling} for $t^\mu_{\phantom{\mu}\nu}$ was chosen to be $l^3/(4\pi G_N)$ instead of $l^3/(16\pi G_N)$.

It is clear from \eqref{g-FG-expansion} that schematically
\begin{eqnarray}
R &\approx& \frac{1}{\Lambda^4} \cdot \Box t + \mathcal{O}\left(\frac{1}{\Lambda^6}\right), \quad \nabla^2 R \approx \frac{1}{\Lambda^4} \cdot \Box^2 t + \mathcal{O}\left(\frac{1}{\Lambda^6}\right), \nonumber\\ R^2 &\approx& \frac{1}{\Lambda^8} \cdot \left(\Box t\right)^2+ \mathcal{O}\left(\frac{1}{\Lambda^{10}}\right)\quad\text{etc.}
\end{eqnarray}
when the boundary metric is flat Minkowski space. Therefore it is clear that only a finite number of geometric counter-terms in $t^\mu_{\phantom{\mu}\nu}$, as defined by \eqref{tFG}, can contribute up to a fixed order in $\Lambda^{-n}$ in the UV expansion. In order to determine all terms in \eqref{t-FG-expansion} up to order $\Lambda^{-6}\, \log \, \Lambda$, the three counterterms (namely the one proportional to $\delta^\mu_{\phantom{\mu}\nu}$, the one proportional to the Einstein tensor and the $\log$ counterterm) are sufficient, except only for the coefficient $b_{4c}$, which can be affected by a non-minimalist $\nabla^2 R$ type of counterterm that may appear without the $\log r$ pre-factor. For the purposes of illustration of the mapping between highly efficient RG flow and classical gravity equations, we will not consider such a possible counterterm here, although our infrared criterion may demand such a counterterm to exist with a precise coefficient, as discussed in the previous subsection. The $\log$ counterterm determines the coefficient $\tilde{b}_{4c}$.

Substituting the Anstaz for the UV expansions \eqref{g-FG-expansion} and \eqref{t-FG-expansion} in the defining relation for $t^\mu_{\phantom{\mu}\nu}$ given by \eqref{t-z4d}, replacing $\Lambda$ by $1/r$ and the coefficients of the UV expansion of $g_{\mu\nu}$ in \eqref{g-FG-expansion} with already determined values as in \eqref{cns}, we can determine the coefficients of the UV expansion of $t^\mu_{\phantom{\mu}\nu}$ in \eqref{t-FG-expansion}. We obtain
\begin{equation}\label{bns}
b_2 = -\frac{1}{4},  \quad  b_{4a} = 0, \quad b_{4b} = \frac{1}{16}, \quad b_{4c} = \frac{1}{64}, \quad \tilde{b}_{4c} = - \frac{1}{128}, \quad{\rm etc.}
\end{equation}
in $d=4$. The relation between $t^\mu_{\phantom{\mu}\nu}$ and ${t^\mu_{\phantom{\mu}\nu}}^\infty$ can be inverted with the coefficients as above. This inverted relation ireads:
\begin{eqnarray}
{t^{\mu}_{\phantom{\mu}\nu}}^\infty &=&{t^{\mu}_{\phantom{\mu}\nu}} + \frac{1}{\Lambda^2} \cdot \frac{1}{4} \cdot \Box t^{\mu}_{\phantom{\mu}\nu}+ \frac{1}{\Lambda^4}\cdot \left(-\frac{1}{16}\cdot \delta^\mu_{\phantom{\mu}\nu}\cdot
{t^\alpha_{\phantom{\alpha}\beta}}
{t^\beta_{\phantom{\beta}\alpha}} + \frac{3}{64}\cdot \Box^2 {t^\mu_{\phantom{\mu}\nu}}\right)
\nonumber\\&&
+\frac{1}{\Lambda^4}\,\log \, \Lambda \,\left(\frac{1}{128}\Box^2 {t^\mu_{\phantom{\mu}\nu}}\right) + \mathcal{O}\left(\frac{1}{\Lambda^6} \log \, \Lambda\right)\label{t-expansion-inverse}\, .
\end{eqnarray}
Finally using the above and the original UV expansion \eqref{t-FG-expansion} with the known values of coefficients as in \eqref{bns} again, we obtain the highly efficient RG flow equation \eqref{t-rg-example}. 

When the inverted UV expansion \eqref{t-expansion-inverse} is substituted in \eqref{g-FG-expansion} with the known values of coefficients as in \eqref{cns}, the UV expansion of $g_{\mu\nu}$ given by \eqref{g-example} as a functional of $t^{\mu}_{\phantom{\mu}\nu}$ instead of ${t^{\mu}_{\phantom{\mu}\nu}}^\infty$ is obtained. We thus derive the highly efficient RG flow equation \eqref{t-rg-example} and the associated $g_{\mu\nu}$ as in \eqref{g-example} from the classical gravity equations. This completes the task of deriving the highly efficient RG flow equations (\ref{t-rg-example}) and the associated effective metric (\ref{g-example}) from Einstein's gravity.

The obvious question is whether we can do the reverse, meaning starting from the highly efficient RG flow \eqref{t-rg-example}, can we go back to the classical gravity equations?

Firstly the highly efficient RG flow equation \eqref{t-rg-example} can be readily solved in the UV expansion and the result is \eqref{t-FG-expansion} with coefficients as in \eqref{bns}. Of course, \eqref{t-rg-example} is a first order equation, so the only integration constant needed is $t^{\mu}_{\phantom{\mu}\nu}$ at $\Lambda= \infty$, which is ${t^{\mu}_{\phantom{\mu}\nu}}^\infty$.

The RG flow equation \eqref{t-rg-example} is constructed in flat Minkowski space, its solution in the UV expansion is given by \eqref{t-FG-expansion} with coefficients as in \eqref{bns}. The question is whether there is a unique $g_{\mu\nu}$ at each $\Lambda$ so that $\nabla_{(\Lambda)\mu} t^{\mu}_{\phantom{\mu}\nu} = 0$ is satisfied for all $\Lambda$, and also such that $g_{\mu\nu}$ takes a state-independent form (meaning that it is a functional of $t^{\mu}_{\phantom{\mu}\nu}$ and $\Lambda$ only in flat Minkowski space background). 

We have already shown that if a RG flow is indeed highly efficient, the associated $g_{\mu\nu}$ must lead to a bulk metric that satisfies classical gravity equations in one higher dimension with full diffeomorphism invariance. Furthermore, the infrared criterion ensures that the map between a highly efficient RG flow and the dual classical gravity theory is unique. Therefore, it is guaranteed that there is a unique $g_{\mu\nu}$ -- when the highly efficient RG flow is given by (\ref{t-rg-example}), and it is given by (\ref{g-example}) as it the RG flow maps to Einstein's gravity.

Practically, one can obtain $g_{\mu\nu}(\Lambda)$ from the highly efficient RG flow (\ref{t-rg-example}) as follows.   We simply exploit the known unique map between highly efficient RG flow and Einstein's gravity. Substituting the Ansatz for the UV expansion of $g_{\mu\nu}$ as given by \eqref{g-FG-expansion} into the defining relation for $t^{\mu}_{\phantom{\mu}\nu}$ given by \eqref{tFG}, and using the known coefficients \eqref{bns} in the UV expansion \eqref{t-FG-expansion} of the $t^{\mu}_{\phantom{\mu}\nu}$ which solves the RG flow equation \eqref{t-rg-example}, we can solve the unknown coefficients in the UV expansion for $g_{\mu\nu}$. We thus obtain \eqref{cns}. It is the same solution for $g_{\mu\nu}$ that can be directly obtained from Einstein's equations. However, here instead of using Einstein's equations we have used the defining relation of $t^{\mu}_{\phantom{\mu}\nu}$ given by \eqref{tFG} to construct $g_{\mu\nu}$ out of $t^{\mu}_{\phantom{\mu}\nu}$. Once again, this is a first order equation and the integration constant is given by the requirement that at $\Lambda = \infty$, $g_{\mu\nu}$ should correspond to the actual background metric of the field theory, namely $\eta_{\mu\nu}$.

\paragraph{Higher derivative gravity:} The entire discussion can also be repeated for higher derivative corrections to Einstein's equations, provided these can be treated perturbatively in powers of the inverse CFT coupling constants as in $\lambda_{(i)}^{-n}$, with $n>0$, signalling systematic departure from infinitely strongly interacting limit. We recall these coupling constants $\lambda_{(i)}$ map to the relative coefficients of the higher derivative corrections (to Einstein's equations) multiplied by appropriate powers of $l$, which are specified by the dimensionless parameters $\alpha'_{(i)}/l^2$. Thus at a fixed order in $\lambda_{(i)}^{-n}$, we need to take into account only a finite number of higher derivative corrections in gravity. These higher derivative corrections should be such that the UV expansions \eqref{g-FG-expansion} and \eqref{t-FG-expansion} of $g_{\mu\nu}$ and $t^{\mu}_{\phantom{\mu}\nu}$ respectively are not modified, although their coefficients are. These corrections can then be expanded systematically in powers of the inverse CFT coupling constants.

\paragraph{Alternative formulation:} An alternative approach to construction of the invertible map between the highly efficient RG flow and the dual classical gravity equations is to use the formulation of the classical gravity equations (in the Fefferman-Graham gauge) as done in \cite{Kuperstein:2013hqa}. 

We first need to invert the relation between $z^{\mu}_{\phantom{\mu}\nu}$ and $t^{\mu}_{\phantom{\mu}\nu}$ as given by \eqref{tFG}. After some simple linear algebra, we get
\begin{equation}
\label{t-z}
z^{\mu}_{\phantom{\mu}\nu} = r^{d-1}\Bigg(t^\mu_{\phantom{\mu}\nu} - \frac{\text{Tr} \, t}{d-1}  \delta^\mu_{\phantom{\mu}\nu}\Bigg) - \frac{2r}{d-2} \Bigg(R^{\mu}_{\phantom{\mu}\nu} - \frac{R}{2(d-1)} \delta^\mu_{\phantom{\mu}\nu}\Bigg)+ ... \, .
\end{equation}
The inverted relation above can be readily substituted in the tensor equation \eqref{tensor-equation} to obtain an equation for the radial evolution of $t^{\mu}_{\phantom{\mu}\nu}$. At first substituting \eqref{t-z} in \eqref{tensor-equation} (for $d>2$), we obtain 
\begin{eqnarray}
\label{EMTensorEoM-Full}
&&
\frac{\partial t^\mu_{\phantom{\mu} \nu}}{\partial r} - \frac{2 r^{2-d}}{d-2} \frac{\partial R^\mu_{\phantom{\mu} \nu}}{\partial r} - 
 \dfrac{r^{d-1}}{2(d-1)}\left( \textrm{Tr}\, t + r^{2-d} R \right) \left( t^\mu_{\phantom{\mu} \nu} - \frac{2 r^{2-d}}{d-2} R^\mu_{\phantom{\mu} \nu} \right) +
\nonumber\\
&& 
\,
 + \frac{1}{d-1}\Bigg( - \frac{\partial \textrm{Tr}\, t}{\partial r} + \dfrac{r^{2-d}}{d-2} \frac{\partial R}{\partial r}+ \frac{ \textrm{Tr}\, t}{r} + \nonumber\\ &&
 \qquad\qquad
 +\frac{r^{d-1}}{2(d-1)} 
 \left( \textrm{Tr} \,t + r^{2-d} R \right) \left( \textrm{Tr}\, t - \dfrac{r^{2-d}}{d-2} R \right) \Bigg) \delta^\mu_{\phantom{\mu} \nu} + \ldots
 = 0 \, .
\end{eqnarray}
Above we have included all terms which are relevant up to third order in derivatives. Furthermore using the identities,
\begin{eqnarray}
\frac{\partial\Gamma^{\mu}_{\nu\rho} }{\partial r}&=& \frac{1}{2}\left(\nabla_\nu z^\mu_{\phantom{\mu}\rho} + \nabla_\rho z^\mu_{\phantom{\mu}\nu}- \nabla^\mu z_{\nu\rho}\right), \\
\frac{\partial R^{\mu}_{\phantom{\mu}\nu\rho\sigma}}{\partial r}
&=& \frac{1}{2} \left(  \nabla_\rho \nabla_\nu z^\mu_{\phantom{\mu}\sigma}  - \nabla_\sigma \nabla_\nu z^\mu_{\phantom{\mu}\rho}- \nabla_\rho \nabla^\mu z_{\nu\sigma} + \nabla_\sigma \nabla^\mu z_{\nu\rho} \right)
\nonumber\\&&  + \frac{1}{2}\left(  R^{\mu}_{\phantom{\mu}\kappa\rho\sigma}
z^\kappa_{\phantom{\mu}\nu} -
R^{\kappa}_{\phantom{\mu}\nu\rho\sigma}
z^\mu_{\phantom{\mu}\kappa} \right) \, , \quad \text{etc.,}
\end{eqnarray}
and iteratively substituting $z^{\mu}_{\phantom{\mu}\nu}$ in terms of $t^{\mu}_{\phantom{\mu}\nu}$ using \eqref{t-z}, we obtain the second equation of the Hamilton-Jacobi like form as schematically given by:
\begin{equation}
\frac{\partial t^\mu_{\phantom{\mu} \nu}}{\partial r} = K^\mu_{\phantom{\mu} \nu}[t^\mu_{\phantom{\mu} \nu}, \nabla t^\mu_{\phantom{\mu} \nu}, R^\mu_{\phantom{\mu}\nu\rho\sigma}, \nabla R^\mu_{\phantom{\mu}\nu\rho\sigma}].
\end{equation}
Crucially terms on the right hand side above has no radial derivative, and thus the above equation is a first order equation. 

Obviously, this equation for the radial evolution of $t^{\mu}_{\phantom{\mu}\nu}$ is not a highly efficient RG flow unless we can eliminate $g_{\mu\nu}$. Indeed one can substitute the Anstaz for the UV expansions of $g_{\mu\nu}$ and $t^{\mu}_{\phantom{\mu}\nu}$ given by \eqref{g-FG-expansion} and \eqref{t-FG-expansion} respectively in \eqref{t-z} and \eqref{EMTensorEoM-Full}, to solve for all the unknown coefficients as well. The equation \eqref{EMTensorEoM-Full} then reduces to the flat space highly efficient RG flow \eqref{t-rg-example} in the UV expansion, where $g_{\mu\nu}$ has disappeared.

The above formulation is not only useful to understand how the highly efficient RG flow looks like when it maps to classical gravity equations in other gauges and to see its symmetries (we will need to use it in Appendices \ref{diffeotransformderivation} and \ref{diffeotransformderivationUV} to derive the action of $(d+1)-$diffeomorphisms on the Fefferman-Graham RG flow), but it is also useful if we want to find the highly efficient RG flow summed over \textit{all orders in $\Lambda^{-1}$}, but up to fixed order in derivatives. The procedure for doing this in the hydrodynamic limit has been found in \cite{Kuperstein:2013hqa}, where $g_{\mu\nu}$ was eliminated to find the RG flow equation to all orders in $\Lambda^{-1}$, up to fixed order in derivatives. This is crucial for fixing the right infrared end point order by order in derivatives in the hydrodynamic derivative expansion. We will complete this task in our forthcoming publication \cite{secondpart}.

\section{Bulk diffeomorphisms and the \textit{lifted Weyl symmetry} of the RG flow}\label{lifted-Weyl-symmetry}

\subsection{The action of bulk diffeomorphisms on the Fefferman-Graham RG flow}\label{diffeo-action}

When the highly efficient RG flow maps to classical gravity equations, bulk diffeomorphisms are the most general transformations which preserve the form of the Ward identity for local conservation of energy and momentum (\ref{WILambda}) in an appropriately redefined background. Here we study the action of general infinitesimal bulk diffeomorphisms which take the RG flow away from the Fefferman-Graham gauge in the corresponding gravity equations.

The $(d+1)-$dimensional metric itself does not appear in the RG flow equation as in \eqref{t-rg-example}. Nevertheless, when the highly efficient RG flow maps to classical gravity equations, there is a unique relation between ADM variables parametrising the $(d+1)-$dimensional metric and $t^\mu_{\phantom{\mu}\nu}$, at any arbitrary scale (identified with the inverse radial coordinate), and for any arbitrary gauge choice on the gravity side. In case of Einstein's gravity, this relation is given via \eqref{Brown-York}, \eqref{extrinsic-curvature}, \eqref{tandT}, \eqref{Tct} and \eqref{handCi}, along with the infrared criterion which determines all gravitational counterterm coefficients. Thus the change of ADM variables under $(d+1)-$diffeomorphisms induce a unique transformation for $t^\mu_{\phantom{\mu}\nu}$ at any arbitrary scale.

As mentioned before, at least in the Fefferman-Graham gauge one can invert the map and obtain the ADM variables of the $(d+1)-$dimensional bulk metric from $t^\mu_{\phantom{\mu}\nu}$. The inversion is possible in any other gauge also if we can decode the gauge fixing of diffeomorphisms in the corresponding gravity equations from the structural property of the RG flow -- we will return to this issue in the next subsection.

Under an arbitrary infinitesimal bulk diffeomorphism transformation: 
\begin{equation}\label{diffeodef}
r= \tilde{r} +\rho(\tilde{r}, \tilde{x}),  \quad x^\mu = \tilde{x}^\mu + \chi^\mu (\tilde{r}, \tilde{x})
\end{equation}
about the Fefferman-Graham gauge, the transformed ADM variables take the form:
\begin{eqnarray}\label{ADMdiffeo}
\tilde{\alpha} &=& \frac{l}{\tilde{r}}\Big(1-\frac{\rho}{\tilde{r}}+ \frac{\partial\rho}{\partial \tilde{r}}\Big), \nonumber\\
\tilde{\beta}^\mu &=& \frac{\partial \chi^\mu}{\partial \tilde{r}} + g^{\mu\nu}   \frac{\partial\rho}{\partial \tilde{x}^\nu}, \nonumber\\
\tilde{\gamma}_{\mu\nu} &=& \frac{l^2}{\tilde{r}^2}\Big(g_{\mu\nu}+ \rho \frac{\partial g_{\mu\nu}}{\partial \tilde{r}}- 2 \,\frac{\rho}{\tilde{r}}\, g_{\mu\nu}+ \mathcal{L}_\chi g_{\mu\nu}\Big).
\end{eqnarray}
Note, for the sake for brevity from now on, we will avoid explicitly writing that the variables on both sides of the equations relating old and new variables after diffeomorphisms are functions of the new coordinates $\tilde{r}$ and $\tilde{x}$ as above. In fact, transforming to new variables while comparing both the new and old variables at the new coordinate point, forms the core of the definition of the diffeomorphism transformation.

Note the equivalent form of (\ref{diffeodef}) is:
\begin{equation}\label{diffeodefnew}
\Lambda= \tilde{\Lambda} +\rho_\Lambda(\tilde{\Lambda}, \tilde{x}),  \quad x^\mu = \tilde{x}^\mu + \chi^\mu (\tilde{r} = \tilde{\Lambda}^{-1}, \tilde{x})
\end{equation}
with $\rho_\Lambda = - \tilde{\Lambda}^2 \rho$. 

The relation between $\gamma_{\mu\nu}$ and the effective metric $g_{\mu\nu}$ of the highly efficient RG flow is unique as discussed before and is given by \eqref{gandgamma}. This relation along with \eqref{ADMdiffeo} readily implies that
\begin{equation}\label{gdiffeo}
\tilde{g}_{\mu\nu} = g_{\mu\nu} - \rho \,\tilde{\Lambda}^2\, \frac{\partial g_{\mu\nu}}{\partial \tilde{\Lambda}}- 2\,\rho\, \tilde{\Lambda} \, g_{\mu\nu}+ \mathcal{L}_\chi g_{\mu\nu},
\end{equation}
after $\tilde{r}$ is identified with $\tilde{\Lambda}^{-1}$. The above form is remarkably simple. - the transformation of the effective metric is just a combination of (i) a translation in scale by $-\rho\, \tilde{\Lambda}^2$, (ii) a Weyl transformation by $\rho \,\tilde{\Lambda}$, and (iii) a $d-$dimensional diffeomorphism by $\chi^\mu$, at any arbitrary scale. Note that this simple form arises if we perform an infinitesimal bulk diffeomorphism transformation \textit{only} about the Fefferman-Graham gauge and not about any arbitrary gauge.

A crucial remark is that \textit{the transformation \eqref{gdiffeo} of the effective metric $g_{\mu\nu}$ under bulk diffeomorphisms is true for any arbitrary classical theory of gravity corresponding to the highly efficient RG flow} as this arises purely from the kinematics of the transformation \eqref{ADMdiffeo} of the ADM variables. Furthermore, \textit{as the transformation \eqref{ADMdiffeo} of ADM variables also uniquely determines the transformation of $t^\mu_{\phantom{\mu}\nu}$, the latter has a hidden universal structure as well, although it depends on the relation of $t^\mu_{\phantom{\mu}\nu}$ to the ADM variables in the specific classical gravity theory corresponding to the RG flow also}.

We do not want the diffeomorphism to change boundary data. In particular, it should leave the boundary metric invariant. Therefore, $\rho$ should have an asymptotic expansion of the form
\begin{equation}
\rho = r^2 \rho_{(2)} + r^3\rho_{(3)} + \cdots\, .
\end{equation}
Similarly, requiring that $\beta$ vanishes asymptotically implies the following asymptotic expansion of $\chi^\mu$:
\begin{equation}
\chi^\mu = r^2 \chi_{(2)}^\mu + \cdots\,.
\end{equation}

Let us now obtain the transformation of $t^\mu_{\phantom{\mu}\nu}$ when the RG flow maps to Einstein's gravity. Firstly, using the defining relation \eqref{extrinsic-curvature} for the extrinsic curvature and \eqref{ADMdiffeo}, we can obtain its transformation which takes the form
\begin{equation}\label{Kdiffeo}
\tilde{K}_{\mu\nu} = K_{\mu\nu} + \rho \, \frac{\partial K_{\mu\nu}}{\partial \tilde{r}} + \mathcal{L}_\chi K_{\mu\nu} + \frac{l}{\tilde{r}}\nabla_\mu\nabla_\nu \rho,
\end{equation}
where $\nabla_\mu$ is the covariant derivative constructed from the effective scale-dependent metric $g_{\mu\nu}$. For intermediate steps in the derivation of the above, please see Appendix \ref{diffeotransformderivation}.

The transformation of $t^\mu_{\phantom{\mu}\nu}$ has to be found in an expansion, either in the UV expansion in $\Lambda^{-1}$ or in the derivative expansion (the latter is an expansion in derivatives of field-theory coordinates only of course). Let us use the derivative expansion for the purpose of illustration.  Assuming that $\rho$ and $\chi^\mu$ start at zeroth order in derivatives, the Einstein tensor counter-term in $t^\mu_{\phantom{\mu}\nu}$ does contribute to the transformation under bulk diffeomorphisms up to this order, but higher order counterterms do not. The result as readily obtained from \eqref{ADMdiffeo} and \eqref{Kdiffeo} for infinitesimal diffeomorphisms about Fefferman-Graham gauge in $d>2$ reads:
\begin{eqnarray}\label{tdiffeo}
\tilde{t}{^\mu_{\phantom{\mu}\nu}}  &=&{t^\mu_{\phantom{\mu}\nu}} + \rho \frac{\partial}{\partial\tilde{r}} {t^\mu_{\phantom{\mu}\nu}}+ d\frac{\rho}{\tilde{r}}{t^\mu_{\phantom{\mu}\nu}} + \mathcal{L}_\chi {t^\mu_{\phantom{\mu}\nu}}+\nonumber\\&&
+\frac{\tilde{r}}{d-2}\Bigg(\nabla^\mu t^\alpha_{\phantom{\alpha}\nu}\nabla_\alpha\rho+\nabla_\nu t^\mu_{\phantom{\mu}\alpha}\nabla^\alpha\rho+t^\mu_{\phantom{\mu}\alpha}\nabla_\nu \nabla^\alpha\rho+t^\alpha_{\phantom{\alpha}\nu}\nabla^\mu \nabla_\alpha\rho -\nonumber\\&&
\qquad\qquad-2\nabla^\alpha t^\mu_{\phantom{\mu}\nu}\nabla_\alpha\rho -t^\mu_{\phantom{\mu}\nu}\nabla^2\rho -\delta^\mu_{\phantom{\mu}\nu}t^\alpha_{\phantom{\alpha}\beta}\nabla_\alpha\nabla^\beta\rho\Bigg)+\mathcal{O}(\nabla^3).
\end{eqnarray} 
It is to be understood that $\nabla$ is the covariant derivative formed from $g$ expressed as a function of the new coordinates. Above all indices have been lowered and raised with the inverse scale-dependent effective metric $g$ and its inverse, respectively. The derivation of the above is presented in Appendix \ref{diffeotransformderivation}. As $g$ explicitly appears in the transformation of $t$, from the point of view of the RG flow equation, we must re-express $g$ in terms of $t$ as in \eqref{g-example} and substitute it in the above equation, to obtain the transformation of $t$ in terms of $t$ alone. This however in the general case can only be done in an UV expansion, meaning in a power series expansion in $r$. Furthermore, we also need to substitute the explicit form of $\partial t/\partial r$ as in \eqref{t-rg-example}, which we also know in the UV expansion in the general case only. In the subsequent publication \cite{secondpart}, we will be able to give an explicit form summing over all orders in $r^n$ in the hydrodynamic limit.

Nevertheless we can construct a double expansion in $r$ and derivatives. We can readily substitute \eqref{t-rg-example} and \eqref{g-example} in \eqref{tdiffeo}, and after replacing $\tilde{r}$ by $\tilde{\Lambda}^{-1}$ we obtain 
\begin{eqnarray}\label{tdiffeoUVderiv}
\tilde{t}{^\mu_{\phantom{\mu}\nu}}  &=&{t^\mu_{\phantom{\mu}\nu}} + \Bigg(-\frac{1}{\tilde{\Lambda^3}}\rho_{(2)}\cdot\frac{1}{2}\cdot \Box{t^\mu_{\phantom{\mu}\nu}}  + d \left(\frac{1}{\tilde{\Lambda}}\rho_{(2)}+\frac{1}{\tilde{\Lambda}^2}\rho_{(3)}+\frac{1}{\tilde{\Lambda}^3}\rho_{(4)}\right){t^\mu_{\phantom{\mu}\nu}} 
\nonumber\\&&+ \frac{1}{\tilde{\Lambda}^2} \mathcal{L}_{\chi_{(2)}} {t^\mu_{\phantom{\mu}\nu}}+\frac{1}{\tilde{\Lambda}^3} \mathcal{L}_{\chi_{(3)}} {t^\mu_{\phantom{\mu}\nu}}+\nonumber\\&&
+\frac{1}{(d-2)\tilde{\Lambda}^3}\Bigg(\eta^{\mu\beta}\,\partial_\beta t^\alpha_{\phantom{\alpha}\nu}\, \partial_\alpha\rho_{(2)}+\partial_\nu t^\mu_{\phantom{\mu}\alpha}\,\partial_\beta\rho_{(2)}\,\eta^{\alpha\beta}+t^\mu_{\phantom{\mu}\alpha}\,\partial_\nu \partial_\beta\rho_{(2)}\,\eta^{\alpha\beta}+t^\alpha_{\phantom{\alpha}\nu}\,\partial_\beta \partial_\alpha\rho_{(2)}\,\eta^{\beta\mu}-\nonumber\\&&
\qquad\qquad\qquad-2\, \partial_\alpha t^\mu_{\phantom{\mu}\nu}\,\partial_\beta\rho_{(2)}\,\eta^{\alpha\beta} -t^\mu_{\phantom{\mu}\nu}\,\partial_\alpha\partial_\beta\rho_{(2)}\, \eta^{\alpha\beta} -\delta^\mu_{\phantom{\mu}\nu}\, t^\alpha_{\phantom{\alpha}\beta}\,\partial_\alpha\partial_\gamma\rho_{(2)}\,\eta^{\beta\gamma}\Bigg)+\nonumber\\&&+\mathcal{O}\left(\frac{1}{\tilde{\Lambda}^4}\right)\Bigg)+\mathcal{O}\left(\partial^3\right),
\end{eqnarray} 
when the field theory lives in flat Minkowski space. Here $\partial_\mu$ is a short form of $\partial/\partial \tilde{x}^\mu$, i.e. the ordinary partial derivative with respect to the new coordinate.

Remarkably, if we look for terms up to first order in derivatives only, we see the same pattern as in case of the transformation of $g_{\mu\nu}$ as in Eq. \eqref{gdiffeo} -- namely it is a combination of (i) a translation in scale by $-\rho\, \tilde{\Lambda}^2$, (ii) a Weyl transformation by $\rho \,\tilde{\Lambda}$, and (iii) a $d-$dimensional diffeomorphism by $\chi^\mu$, at any arbitrary scale. This simple intuitive form up to first order in derivatives holds for infinitesimal diffeomorphism transformations about Fefferman-Graham gauge only. 

In order to find the new RG flow equation, we need to apply \eqref{diffeodefnew} with the transformation \eqref{tdiffeo} for $t^\mu_{\phantom{\mu}\nu}$ to the Fefferman-Graham RG flow equation, namely \eqref{t-rg-example}. It is clear then that the new RG flow equation, corresponding to the new gauge, depends on $\rho$ and $\chi^\mu$ explicitly. 

The crucial point is that the equations that determine $\rho$ and $\chi^\mu$ in terms of new gauge conditions which specify $\alpha$ and $\beta^\mu$ are first order in $r-$derivative as evident from (\ref{ADMdiffeo}). Therefore, requiring that they do not change the boundary metric, meaning that they disappear in the UV keeping the field theory background metric unchanged, determine them uniquely. Thus there exist unique $\rho$ and $\chi^\mu$ which transforms the Fefferman-Graham gauge to the new gauge provided that the transformation is trivial in the UV \footnote{For an explicit example see \cite{Gupta:2008th}, where the change of coordinates from the Fefferman-Graham gauge to Eddington-Finkelstein gauge for a special class of solutions has been worked out.}. Therefore, if we can understand to which gauge fixing in classical gravity the new RG flow equation corresponds to, we can simply apply the reverse transformation and bring the new equation into Fefferman-Graham gauge where $t^\mu_{\phantom{\mu}\nu}$ and $\Lambda$ only appear explicitly. Thus we are led to the problem of understanding the link between the structural property of the RG flow equation and the corresponding gauge fixing on the gravity side. We will solve this problem in Section \ref{decipher}. 

As of now, we conclude that the new RG flow equations obtained from arbitrary bulk diffeomorphism transformations are also highly efficient, in the sense that they preserve the form of the Ward identity (\ref{WILambda}) in an appropriately redefined background metric at each scale, although they are not state-independent explicitly. The state-dependence enters through the auxiliary non-dynamical variables $\alpha$ and $\beta^\mu$ only. With an appropriate diffeomorphism transformation discussed here, they can be brought back to the form where they depend only on $t^\mu_{\phantom{\mu}\nu}$ and $\Lambda$ explicitly. In that case, the highly efficient RG flow corresponds to the classical gravity equations in the Fefferman-Graham gauge.

\subsection{The lifted Weyl symmetry}\label{lifted-Weyl}

A CFT on a curved background is Weyl invariant up to the Weyl anomaly. By the holographic duality, Weyl transformations at the boundary can be lifted to specific diffeomorphisms in the $(d+1)-$dimensional geometry. These specific diffeomorphisms actually depend on the geometry which solves classical gravity equations and hence on the dual field theory state, however their dependence on the $(d+1)-$dimensional metric has a form which is independent of the geometry and hence of the dual state.  

These diffeomorphisms are actually the set of diffeomorphisms which preserve the asymptotic AdS nature of the $(d+1)-$dimensional metric but yet do not vanish at the boundary, where they manifest as pure Weyl transformations. Furthermore, these are the residual gauge symmetries of the Fefferman-Graham gauge. 

Clearly, the general transformation of ADM variables \eqref{ADMdiffeo} away from the Fefferman-Graham gauge may be performed in such a way that the Fefferman-Graham gauge is preserved, meaning that $\tilde{\alpha}$ remains $l/\tilde{r}$ and $\tilde{\beta}^\mu$ still vanishes, provided in the diffeomorphisms (\ref{diffeodef}),
\begin{equation}\label{PBH}
\rho = \tilde{r}\,\delta\sigma(\tilde{x}), \quad \chi^\mu = -\int_0^{\tilde{r} } {\rm d}\hat{r} \,\hat{r} \, g^{\mu\nu}(\hat{r}, \tilde{x}) \frac{\partial \delta\sigma(\tilde{x})}{\partial \tilde{x}^\nu}.
\end{equation}
It is clear that $\chi^\mu$ depends on the specific geometry concerned through $g_{\mu\nu}$, but its dependence on $g_{\mu\nu}$ takes the above state-independent form. These are specific forms of so-called Penrose-Brown-Henneaux (PBH) transformations \cite{Penrose,Brown:1986nw,Schwimmer:2000cu}. When rewritten after replacing the radial coordinate by the inverse of the scale, the above transformations take the form: 
\begin{equation}\label{PBHLambda}
\rho_\Lambda = - \tilde{\Lambda}\,\delta\sigma(\tilde{x}), \quad \chi^\mu = -\int_{\tilde{\Lambda}}^\infty d\hat{\Lambda} \, \frac{1}{\hat{\Lambda}^3}g^{\mu\nu}(\hat{\Lambda}, \tilde{x}) \frac{\partial \delta\sigma(\tilde{x})}{\partial \tilde{x}^\nu},
\end{equation}

The transformations depend on one parameter only, namely $\delta\sigma(x)$ which is a function of the field theory coordinates, and is nothing but the Weyl transformation parameter in the UV.

It is also to be noted that although these transformations preserve the Fefferman-Graham gauge, they are not isometries in a generic geometry. Indeed, the effective metric $g_{\mu\nu}$ still transforms non-trivially according to \eqref{gdiffeo}. It is also easy to see explicitly from here that at the boundary, meaning at $\Lambda=\infty$, the transformation reduces to a Weyl rescaling of the background metric by $\delta\sigma$. In particular, if the boundary metric is conformally flat, i.e. $g_{\mu\nu}(r=0) =\eta_{\mu\nu}e^{2\sigma(x)}$, the new boundary metric is also conformally flat but, with $\sigma$ is replaced by $\sigma +\delta\sigma$. In this case, the asymptotic forms of $\rho$, $\chi^\mu$ are
\begin{equation}
\rho = \frac{1}{\tilde{\Lambda}} \, \delta\sigma, \quad \chi^\mu = - \frac{1}{2\tilde{\Lambda}^2} e^{-2\sigma}\eta^{\mu\nu}\, \frac{\partial}{\partial{\tilde{x}^\nu}}\delta\sigma + \mathcal{O}\left(\frac{1}{\tilde{\Lambda}^3}\right).
\end{equation}
The asymptotic form of the transformation of $g_{\mu\nu}$ according to \eqref{gdiffeo} is then
\begin{equation}
\tilde{g}_{\mu\nu} = g_{\mu\nu} - 2\, \delta\sigma\,  g_{\mu\nu} + 2\, \frac{1}{\tilde{\Lambda}^2}\, \delta\sigma\,g_{(2)\mu\nu} - \frac{1}{2\tilde{\Lambda}^2}\, \mathcal{L}_{e^{-2\sigma}\eta^{\alpha\beta}\,\partial_\beta\delta\sigma}\, (\eta_{\mu\nu}e^{2\sigma})+ \mathcal{O}\left(\frac{1}{\tilde{\Lambda}^3}\right).
\end{equation}
It is clear that the leading term in the transformation above is a Weyl transformation by $\delta\sigma$ as claimed before. It is worth recalling that all variables on both sides of the above equation  are functions of $\tilde{\Lambda}$, the new scale, and of $\tilde{x}$, the new field theory coordinates. Moreover, $\partial_\beta$ stands for $\partial/\partial\tilde{x}^\beta$, the partial derivative with respect to the new coordinate. In the above formula, $g_{(2)\mu\nu}$, which appears in one of the  the sub-leading terms in the asymptotic expansion of $g_{\mu\nu}$ (and arises from $(\partial/\partial \Lambda)g$), can be obtained directly from Einstein's equations. As derived in Appendix \ref{hergCF}, its explicit form reads
\begin{equation}\label{g2text}
g_{(2)\mu\nu} = - (\partial_\mu \sigma) (\partial_\nu\sigma) + \partial_\mu\partial_\nu\sigma + \frac{1}{2}\eta_{\mu\nu} \eta^{\alpha\beta}(\partial_\alpha \sigma) (\partial_\beta \sigma).
\end{equation}
Thus it a functional of the background metric.

Let us demonstrate that these transformations \eqref{PBHLambda} along with the transformation of $t^\mu_{\phantom{\mu}\nu}$ given by  \eqref{tdiffeo} are a symmetry of the highly efficient RG flow equation corresponding to the Fefferman-Graham gauge (note that the effective metric $g$ never enters this equation so its transformation is irrelevant for the symmetry). In order to recognise the symmetry, we need to generalise the RG flow equation \eqref{t-rg-example} from the flat Minkowski space background to a (non-dynamical) conformally flat space background $\eta_{\mu\nu}e^{2\sigma(x)}$. We will call this the \textit{lifted Weyl symmetry}.

The derivation of the highly efficient RG flow in the background of conformally flat space $\eta_{\mu\nu} e^{2\sigma}$ which maps to Einstein's gravity equations can be obtained following the methodology of Section \ref{calculating-map}. However, the calculations are highly involved and can be simplified with the help of some tricks and useful identities. The derivation is presented in Appendix \ref{hergCF}. The result in $d=4$ reads:
\begin{eqnarray}\label{t-rg-example-CF}
\frac{\partial}{\partial \Lambda} t^\mu_{\phantom{\mu}\nu}(\Lambda) &=& -\frac{1}{\Lambda^3} \Big(2{t^{\mu}_{(2)\nu}}^* + \nonumber\\&&\qquad\qquad
+\frac{1}{2}\Big(-e^{-2\sigma}\eta^{\alpha\beta}\,\nabla_{\alpha}^{(\sigma)}\nabla_{\beta}^{(\sigma)}\,\left(t^\mu_{\phantom{\mu}\nu} (\Lambda)-{t^\mu_{(0)\nu}}^* \right) 
+ \nonumber\\&&\qquad\qquad\qquad
+4 \left(t^\mu_{\phantom{\mu}\beta}(\Lambda) -{t^\mu_{(0)\beta}}^* \right)\,e^{-2\sigma}\eta^{\beta\gamma}\, g_{(2)\gamma\nu} 
+ \nonumber\\&&\qquad\qquad\qquad+ 6\,\delta^\mu_{\phantom{\mu}\nu} \, \left(t^\gamma_{\phantom{\gamma}\beta}(\Lambda) -{t^\gamma_{(0)\beta}}^* \right)\, e^{-2\sigma}\eta^{\beta\delta}\,g_{(2)\gamma\delta}\Big)+ \nonumber\\&&\qquad+ t^\mu_{(2a)\nu}\Big) + 2 \frac{1}{\Lambda^3}\, \log \, \Lambda \, t^\mu_{(2a)\nu} +\mathcal{O}\left(\frac{1}{\Lambda^5} \log \, \Lambda\right).
\end{eqnarray}
Here, $\nabla^{(\sigma)}$ denotes the covariant derivative constructed from the background metric $\eta_{\mu\nu}e^{2\sigma}$. The variables $t{^\mu_{(0)\nu}}^* $, $t{^{\mu}_{(2)\nu}}^*$ and $t{^\mu_{(2a)\nu}}^*$ are independent of $\Lambda$ and functionals of the background metric $\eta_{\mu\nu}e^{2\sigma}$, just like $g_{(2)\mu\nu}$ (which is given by \eqref{g2text}), and they vanish when $\sigma$ goes to zero, i.e. when the background metric reduces to flat Minkowski space. These functions are given explicitly in Appendix \ref{hergCF}. In that case \eqref{t-rg-example-CF} reduces to the flat Minkowski space highly efficient RG flow equation \eqref{t-rg-example} as evident from the leading term. The background dependent term at order $(1/\Lambda^3) \log\Lambda$ appearing in  \eqref{t-rg-example-CF} is completely determined by the conformal anomaly.

The transformation \eqref{PBH} preserves the Fefferman-Graham gauge and also the form of the boundary metric $\eta_{\mu\nu}e^{2\sigma}$, although replacing $\sigma$ by $\sigma +\delta\sigma$. Since the Fefferman-Graham gauge is preserved, the relation between $t$ and $g$ given by \eqref{t-z4d} also remains invariant under the transformation. This then readily implies that the resultant highly efficient RG flow equation \eqref{t-rg-example-CF} should remain the same after the transformation, but with $\sigma$ replaced by $\sigma+\delta\sigma$. Therefore  \eqref{PBH} should be a symmetry of the transformation, provided we use \eqref{t-z4d} to find the transformation of $t^\mu_{\phantom{\mu}\nu}$ from the transformation of $g_{\mu\nu}$, the latter being given by \eqref{gdiffeo} and \eqref{PBH}. This transformation of $t^\mu_{\phantom{\mu}\nu}$ reads (please refer to Appendix \ref{diffeotransformderivationUV} for more details):
\begin{eqnarray}\label{tmunudiffeoCFLambda}
\tilde{t}{^\mu_{\phantom{\mu}\nu}}  &=&{t^\mu_{\phantom{\mu}\nu}} + \Bigg(\frac{1}{\tilde{\Lambda}^2} e^{-2\sigma}\,\delta\sigma\, \frac{1}{2}\Big(-\eta^{\alpha\beta}\,\nabla_{\alpha}^{(\sigma)}\nabla_{\beta}^{(\sigma)} t^\mu_{\phantom{\mu}\nu}
+4 t^\mu_{\phantom{\mu}\beta} \,\eta^{\beta\gamma}\, g_{(2)\gamma\nu} 
+  6\delta^\mu_{\phantom{\mu}\nu} \, t^\gamma_{\phantom{\gamma}\beta}\,\eta^{\beta\delta}\,g_{(2)\gamma\delta}\Big)\nonumber\\&& + 4\,\delta\sigma \,{t^\mu_{\phantom{\mu}\nu}} - \frac{1}{2}\frac{1}{\tilde{\Lambda}^2} \mathcal{L}_{e^{-2\sigma}\eta^{\alpha\beta}\partial_\beta\delta\sigma}\,\, {t^\mu_{\phantom{\mu}\nu}}+\nonumber\\&&
+\frac{e^{-2\sigma}}{2\tilde{\Lambda}^2 }\Bigg(\eta^{\mu\beta}\partial_\beta t^\alpha_{\phantom{\alpha}\nu}\,\partial_\alpha\delta\sigma+\partial_\nu t^\mu_{\phantom{\mu}\alpha}\,\partial_\beta\delta\sigma\,\eta^{\alpha\beta}+t^\mu_{\phantom{\mu}\alpha}\,\partial_\nu \partial_\beta\delta\sigma\,\eta^{\alpha\beta}+t^\alpha_{\phantom{\alpha}\nu}\,\partial_\beta \partial_\alpha\delta\sigma\,\eta^{\beta\mu}-\nonumber\\&&
\qquad\qquad\qquad-2\partial_\alpha t^\mu_{\phantom{\mu}\nu}\,\partial_\beta\delta\sigma\,\eta^{\alpha\beta} -t^\mu_{\phantom{\mu}\nu}\,\partial_\alpha\partial_\beta\delta\sigma\,\eta^{\alpha\beta} -\delta^\mu_{\phantom{\mu}\nu}t^\alpha_{\phantom{\alpha}\beta}\,\partial_\alpha\partial_\gamma\delta\sigma\,\eta^{\beta\gamma}\Bigg)+\nonumber\\&&+\mathcal{O}\left(\frac{1}{\tilde{\Lambda}^4}\right)\Bigg)+\mathcal{O}(\partial^3),
\end{eqnarray} 
up to given orders in derivatives and $\Lambda^{-1}$. The highly efficient RG flow equation \eqref{t-rg-example-CF} in conformally flat space background is invariant under the transformations given by \eqref{PBHLambda} and the above equation, up to the orders in $\Lambda^{-1}$ shown when restricted to second order in derivatives.

The transformation of $t^\mu_{\phantom{\mu}\nu}$ is complicated as evident from \eqref{tmunudiffeoCFLambda}, and in the general case it can only be found in an asymptotic expansion in a specific theory of classical gravity. Nevertheless, it is unique and can be deduced from the transformation of $g_{\mu\nu}$ as given by \eqref{gdiffeo} and \eqref{PBH}, and the latter is simple, exact to all orders in the UV expansion, and also holds for an arbitrary classical theory of gravity. Therefore \eqref{gdiffeo} and \eqref{PBH} should be thought of as more fundamental in defining the lifted Weyl symmetry of the highly efficient RG flow corresponding to the Fefferman-Graham gauge fixing of the dual classical gravity equations. 

In the following subsection, we will find the lifted Weyl symmetry of the highly efficient RG flow corresponding to an arbitrary gauge fixing of diffeomorphism symmetry of the dual classical gravity equations.

\subsection{Deciphering the gauge fixing of diffeomorphisms from the lifted Weyl symmetry}\label{decipher}

We will prove here that every choice of gauge fixing of the $(d+1)-$diffeomorphism symmetry in the corresponding classical gravity equations induces a unique lifted Weyl symmetry in the dual highly efficient RG flow that reduces to Weyl transformations in the UV. The inverse statement is also true, i.e. the gauge fixing of the $(d+1)-$diffeomorphism symmetry in the classical gravity equations can be completely deciphered from the lifted Weyl symmetry of the corresponding highly efficient RG flow. The latter can also be defined in a manner which is also completely independent of the classical gravity theory to which the highly efficient RG flow corresponds to. \textit{Therefore, the lifted Weyl symmetry is an universal feature of highly efficient RG flows,} if indeed as argued in Section \ref{Intro2} the defining property of highly efficient RG flow leads to it being a rephrasing of a classical gravity theory will full diffeomorphism invariance.

Firstly, let us choose a gauge which is infinitesimally different from the Fefferman-Graham one. This can be specified by the gauge-fixing conditions which determine $\alpha$ and $\beta^\mu$ uniquely. From the  first order equations \eqref{ADMdiffeo} relating the new  $\alpha$ and $\beta^\mu$ to those in the case of Fefferman-Graham gauge, we can then determine the infinitesimal diffeomorphism transformations, meaning $\rho$ and $\chi^\mu$, which take us from the Fefferman-Graham gauge to the new gauge. Since in \eqref{ADMdiffeo}, we have only first order radial derivatives of $\rho$ and $\chi^\mu$ given the boundary conditions that $\rho= \mathcal{O}(r^2)$ and $\chi^\mu = \mathcal{O}(r^2)$ so that they do not modify boundary (UV) data, we obtain unique solutions for $\rho$ and $\chi^\mu$. Let us call this unique transformation $\mathcal{G}$.

The lifted Weyl symmetry transformation corresponding to a Weyl transformation by $\delta\sigma$ in the UV, and under which the new highly efficient RG flow equation is invariant reads $\mathcal{G}\mathcal{P}^{\delta\sigma}\mathcal{G}^{-1}$, where $\mathcal{P}^{\delta\sigma}$ corresponds to the PBH transformation \eqref{PBH}. This can be verified as follows: Firstly, with $\mathcal{G}^{-1}$ we can transform from the new gauge back to Fefferman-Graham gauge. Note that this does not modify the UV data and the background metric. Then with $\mathcal{P}^{\delta\sigma}$, we can perform the unique diffeomorphism transformation that is non-vanishing at the boundary ($\Lambda = \infty$) modifying the UV data and the background metric by a Weyl transformation with parameter $\delta\sigma$, while keeping the Fefferman-Graham highly efficient RG flow equations (\ref{t-rg-example-CF}) invariant. Finally, we can go back to the new gauge by the unique transformation $\mathcal{G}$ (which itself does not modify the UV data). In total, we have arrived at the new gauge, but with UV data and background metric Weyl transformed by $\delta\sigma$. Clearly $\mathcal{G}\mathcal{P}^{\delta\sigma}\mathcal{G}^{-1}$ should be the new and unique transformation under which the new highly efficient RG flow equation in conformally flat metric background is invariant, since after the transformation, the RG flow corresponds to the same gauge fixing of diffeomorphisms of the dual classical gravity equations, but with the UV data and the conformally flat background metric are Weyl transformed by $\delta\sigma$. 

The transformation of $g_{\mu\nu}$ then follows from \eqref{gdiffeo}, which in turn results in a unique transformation of $t^\mu_{\phantom{\mu}\nu}$ as discussed in the previous subsections. Recall $g_{\mu\nu}(\Lambda)$ is unique in a highly efficient RG flow, and its transformation also induces a unique transformation of $t^\mu_{\phantom{\mu}\nu}$ because the relation between the two is preserved by the lifted Weyl symmetry. Nevertheless, the more fundamental definition of the lifted Weyl symmetry is via the transformation of $g_{\mu\nu}$, as it takes the same form in any classical theory of gravity. From the transformation of $g_{\mu\nu}$ one can read off the diffeomorphism $\mathcal{G}$ that relates the Fefferman-Graham gauge to the new gauge, given that we can find the new lifted Weyl symmetry $\mathcal{G}\mathcal{P}^{\delta\sigma}\mathcal{G}^{-1}$ and we know $\mathcal{P}^{\delta\sigma}$ explicitly. Thus from the lifted Weyl symmetry we can decipher the corresponding gauge fixing of diffeomorphisms in the dual classical gravity equations. This correspondence between the symmetry of the highly efficient RG flow and the gauge fixing in the corresponding gravity equations is independent of the classical gravity theory to which the highly efficient RG flows map to, as the transformation of $g_{\mu\nu}$ is independent of the classical gravity theory.

%\section{The \textit{exact asymptotic hydrodynamic expansion} and its coarse graining}\label{hydro}
%\section{An explicit construction of a highly efficient RG flow and the dual gravity equations}\label{HERG-example}
\section{Conclusions}\label{conclusions}

In this paper, we have shown that all diffeomorphism invariant pure classical gravity theories can be recast as first order highly efficient RG flows. Furthermore, we have claimed that by imposing an appropriate infrared criterion on the latter, we can also reproduce the right UV data which lead to absence of naked singularities in the solutions of the dual gravity theory. We will establish this claim with explicit examples in the second part of our work \cite{secondpart}. 

We have not yet shown how a precise construction in the field theory involving a specific procedure of coarse-graining can generate such highly efficient RG flows in strongly interacting large $N$ quantum field theories. This will be the main object of study of the second part of this work \cite{secondpart} as well.

Although we have focused on the pure gravity sector here, it is not hard to see that our approach extends beyond the case of pure gravity. Consider the case when we have an additional scalar single-trace operator $O$. In this case, the highly efficient RG flow equations which should reproduce the dual gravity equations, will assume the form:
\begin{eqnarray}
\frac{\partial}{\partial \Lambda}t^\mu_{\phantom{\mu}\nu}(\Lambda) &=& F^\mu_{1\nu}[t^\mu_{\phantom{\mu}\nu}(\Lambda), O(\Lambda),\Lambda ], \nonumber\\
\frac{\partial}{\partial \Lambda}O(\Lambda) &=& F_2[t^\mu_{\phantom{\mu}\nu}(\Lambda), O(\Lambda),\Lambda ].
\end{eqnarray}
The scale-dependent Ward identity that must be preserved at all scales is:
\begin{equation}\label{WILambdanew}
\nabla_{(\Lambda)\mu}t^\mu_{\phantom{\mu}\nu}(\Lambda) = O(\Lambda)\, \partial_\mu J(\Lambda),
\end{equation}
where $g_{\mu\nu}[t^\mu_{\phantom{\mu}\nu}(\Lambda), O(\Lambda),\Lambda ]$ is an appropriate scale-dependent background metric and $J[t^\mu_{\phantom{\mu}\nu}(\Lambda), O(\Lambda),\Lambda ]$  is an appropriate scale-dependent external source that are functionals of the effective single-trace operators. With an extra propagating degree of freedom, a Ward identity of the above formed can be preserved along the scale evolution only if there is an underlying diffeomorphism invariance, such that $g_{\mu\nu}(\Lambda)$ and $J(\Lambda)$ lead to a bulk metric and a bulk scalar field in the Fefferman-Graham gauge, which solve the equations of a classical gravity theory coupled to a scalar field with full $(d+1)-$diffeomorphism invariance. The arguments supporting this assertion will mimic those given in Section \ref{Intro2}. Nevertheless, it is not easy to  formulate the right infrared criterion. We will comment more on this in the second part of the work.

It will be particularly interesting to see how higher spin gravities can be reformulated as highly efficient RG flows. Such theories of gravity have a larger group of gauge symmetries which also include standard diffeomorphisms. We need to understand how these gauge symmetries in $(d+1)-$dimensions can be ensured by $d-$dimensional Ward identities which generalise (\ref{WILambda}) in order to apply our approach to this problem. The insights obtained from \cite{Koch:2010cy,Douglas:2010rc,Gopakumar:2011ev,Sachs:2013pca,Leigh:2014qca,Mintun:2014gua} should be highly valuable in this regard. This development is also worth taking because the dual field theories are often solvable particularly for the case of $d=2$. We can then hope to learn a lot about how holographic correspondence leads to alternative formulations of quantum field theories.

Finally, we emphasise again that our approach for reconstructing the holographic correspondence as a highly efficient RG flow should also work when the solutions of the classical gravity theory do not have a well defined asymptotic region, so that we need to impose a UV cut-off to make sense of the correspondence. As in our approach all UV data can be determined through our infrared criterion, we do not need an asymptotic UV region on the gravity side to have a precise map between the theory of gravity and effective dual field theory at a given scale. Nevertheless, it will be interesting to work out such an example explicitly.

In the second part of our work, we will comment more on possible intersections of our approach with other works in the literature. We will also discuss how our approach can be used to construct new non-perturbative frameworks for theories like Quantum Chromodynamics (QCD) and some open questions.

\acknowledgments 

We thank B. P. Dolan for collaboration during the initial stages of this project. N.B. and A.M. acknowledge the ``Research in Groups'' grant sponsored by ESPRC of UK and managed by ICMS, Edinburgh, which kickstarted their mutual collaboration. A.M. thanks ICMS and Heriot-Watt University for hospitality at Edinburgh, where a large portion of the work in this project has been done. The research of A.M. involving a large part of this work has also been supported by the LABEX P2IO, the ANR contract 05-BLAN- NT09-573739, the ERC Advanced Grant 226371 and the ITN programme PITN-GA-2009-237920. Presently, the research of A.M. is supported in part by European Union's Seventh Framework Programme under grant
agreements (FP7-REGPOT-2012-2013-1) no 316165, the EU-Greece
program ``Thales" MIS 375734, and is also co-financed by the European
Union (European Social Fund, ESF) and Greek national funds through the Operational
Program ``Education and Lifelong Learning" of the National Strategic Reference
Framework (NSRF) under ``Funding of proposals that have received a positive
evaluation in the 3rd and 4th Call of ERC Grant Schemes". S.K. is supported in part by the ANR grant 08-JCJC-0001-0 and the ERC Starting Grants 240210-String-QCD-BH and 259133-ObservableString. We are dedicating this work as a tribute to the occasion of the 100th anniversary of General Relativity.

\appendix
\section{Derivation of transformations under bulk diffeomorphisms in derivative expansion}\label{diffeotransformderivation}
We first note that the transformation of the induced metric $\gamma_{\mu\nu}$ as given by \eqref{ADMdiffeo} can be written compactly in the form:
\begin{equation}\label{simplegammadiffeo}
\tilde{\gamma}_{\mu\nu} = \gamma_{\mu\nu} + \rho \frac{\partial \gamma_{\mu\nu}}{\partial \tilde{r}} + \mathcal{L}_{\chi}\gamma_{\mu\nu}.
\end{equation}
One can readily prove the following result
\begin{equation}
\frac{\partial}{\partial r} \mathcal{L}_\chi A^{\mu_1 ...\mu_n}_{\nu_1 .... \nu_n} = \mathcal{L}_\chi \frac{\partial}{\partial r}A^{\mu_1 ...\mu_n}_{\nu_1 .... \nu_n}+\mathcal{L}_{\frac{\partial\chi}{\partial r}} A^{\mu_1 ...\mu_n}_{\nu_1 .... \nu_n},
\end{equation}
for an arbitrary tensor $ A^{\mu_1 ...\mu_n}_{\nu_1 .... \nu_n}$. With the above rules, one can substitute \eqref{ADMdiffeo} in  the definition of the extrinsic curvature \eqref{extrinsic-curvature}, we readily obtain the transformation of the latter as in \eqref{Kdiffeo}.

It is also not hard to see that the transformation of the inverse induced metric is given by
\begin{equation}\label{simpleinvgammadiffeo}
\tilde{\gamma}^{\mu\nu} = \gamma^{\mu\nu} + \rho \frac{\partial \gamma^{\mu\nu}}{\partial \tilde{r}} + \mathcal{L}_{\chi}\gamma^{\mu\nu}.
\end{equation}

Let us define the bare energy-momentum tensor as ${t^\mu_{\phantom{\mu}\nu}}^{\rm bare}$ as
\begin{equation}
{t^\mu_{\phantom{\mu}\nu}}^{\rm bare} = -2\frac{l}{r^d} \, \gamma^{\mu\rho}\left(K_{\rho\nu}- K \gamma_{\rho\nu}\right).
\end{equation}
in case of Einstein's gravity in a arbitrary $d$. The transformation of ${t^\mu_{\phantom{\mu}\nu}}^{\rm bare}$ is then as follows
\begin{equation}\label{tmunubarediffeo}
\tilde{t}{^\mu_{\phantom{\mu}\nu}}^{\rm bare}  ={t^\mu_{\phantom{\mu}\nu}}^{\rm bare} + \rho \frac{\partial}{\partial\tilde{r}} {t^\mu_{\phantom{\mu}\nu}}^{\rm bare}+ d\frac{\rho}{\tilde{r}}{t^\mu_{\phantom{\mu}\nu}}^{\rm bare} + \mathcal{L}_\chi {t^\mu_{\phantom{\mu}\nu}}^{\rm bare} - \frac{2}{\tilde{r}^{d-1}}\left(\nabla^\mu \nabla_\nu \rho - \nabla^2 \rho \delta^\mu_{\phantom{\mu}\nu}\right).
\end{equation}
We readily see that the last term after transformation diverges at $r=0$ (at the UV) if $\rho$ vanishes as $r$ asymptotically (which is allowed as it does not shift the boundary away from $r=0$)-- this must be cancelled by the transformation of the counterterm. We will see this is indeed the case. Above all indices have been lowered/raised using $g$/its inverse and $\nabla$ is the covariant derivative constructed from $g$.

The first counterterm required to cancel the on-shell volume divergence in ${t^\mu_{\phantom{\mu}\nu}}^{\rm bare}$ is
\begin{equation}
{t^\mu_{\phantom{\mu}\nu}}^{\rm ct(1)}= - 2 (d-1) \frac{1}{r^d}\delta^\mu_{\phantom{\mu}\nu}.
\end{equation}
This does not transform under bulk diffeomorphism at all, simply because in the new coordinates we have to replace $r$ by $\tilde{r}$ (note we are comparing ${t^\mu_{\phantom{\mu}\nu}}^{\rm ct(1)}(\tilde{r}, \tilde{x})$ with $\tilde{t}{^\mu_{\phantom{\mu}\nu}}^{\rm ct(1)}(\tilde{r}, \tilde{x})$ and they are verily the same as $\delta^\mu_{\phantom{\mu}\nu}$ is invariant). Nevertheless, $\mathcal{L}_\chi {t^\mu_{\phantom{\mu}\nu}}^{\rm ct(1)} = 0$ and $(\rho(\partial/\partial r) + d(\rho/r)) {t^\mu_{\phantom{\mu}\nu}}^{\rm ct(1)} = 0$, so we can as well write for later convenience that
\begin{equation}\label{tmunuct1diffeo}
\tilde{t}{^\mu_{\phantom{\mu}\nu}}^{\rm ct(1)}  ={t^\mu_{\phantom{\mu}\nu}}^{\rm ct(1)} + \rho \frac{\partial}{\partial\tilde{r}} {t^\mu_{\phantom{\mu}\nu}}^{\rm ct(1)}+ d\frac{\rho}{\tilde{r}}{t^\mu_{\phantom{\mu}\nu}}^{\rm ct(1)} + \mathcal{L}_\chi {t^\mu_{\phantom{\mu}\nu}}^{\rm ct(1)} .
\end{equation}

The second counterterm required to cancel the subleading on-shell divergence in ${t^\mu_{\phantom{\mu}\nu}}^{\rm bare}$ is
\begin{equation}
{t^\mu_{\phantom{\mu}\nu}}^{\rm ct(2)}= \frac{2}{d-2} \frac{l^2}{r^d}\left(R^\mu_{\phantom{\mu}\nu}[\gamma] - \frac{1}{2}R[\gamma]\delta^\mu_{\phantom{\mu}\nu}\right).
\end{equation}
Identically, for any arbitrary transformation
\begin{eqnarray}
\delta R_{\nu\sigma}[\gamma] = \frac{1}{2}\nabla_\mu\left(\nabla_\nu \gamma^{\mu\rho}\delta\gamma_{\rho\sigma} + \nabla_\sigma \gamma^{\mu\rho}\delta\gamma_{\rho\nu}-\nabla_\rho \gamma^{\mu\rho}\delta\gamma_{\nu\sigma}\right) - \frac{1}{2}\nabla_\sigma\nabla_\nu \gamma^{\alpha\beta}\delta\gamma_{\alpha\beta}.
\end{eqnarray}
To find the transformation under bulk diffeomorphism we simply need to substitute \eqref{simplegammadiffeo} above. Employing in addition the vector Einstein equation \eqref{vector-equation} and after some further rearrangements, we get
\begin{eqnarray}
\delta R_{\nu\sigma}[\gamma] &=& \rho \frac{\partial}{\partial \tilde{r}}R_{\nu\sigma}[\gamma] + \mathcal{L}_\chi R_{\nu\sigma}[\gamma]+\nonumber\\&&
+\frac{1}{2}\Bigg(\nabla_\nu z^\mu_{\phantom{\mu}\sigma}\nabla_\mu \rho +\nabla_\sigma z^\mu_{\phantom{\mu}\nu}\nabla_\mu \rho + z^\mu_{\phantom{\mu}\sigma}\nabla_\nu\nabla_\mu \rho +z^\mu_{\phantom{\mu}\nu}\nabla_\sigma\nabla_\mu \rho-\nonumber\\&&
\qquad - 2\nabla^\mu z_{\nu\sigma}\nabla_\mu\rho - z_{\nu\sigma}\nabla^2 \rho - {\rm Tr}\, z\, \nabla_\nu\nabla_\sigma \rho\Bigg)+
\nonumber\\&&
+ \frac{1}{\tilde{r}}(d-2) \nabla_\nu\nabla_\sigma \rho + \frac{1}{\tilde{r}} \nabla^2\rho \, g_{\nu\sigma}.
\end{eqnarray}
Above $z^\mu_{\phantom{\mu}\nu}$ is as defined in \eqref{z} and $z_{\mu\nu} = g_{\mu\rho}z^\rho_{\phantom{\rho}\nu}$. This implies that
\begin{eqnarray}\label{tmunuct2diffeo}
\tilde{t}{^\mu_{\phantom{\mu}\nu}}^{\rm ct(2)}  &=&{t^\mu_{\phantom{\mu}\nu}}^{\rm ct(2)} + \rho \frac{\partial}{\partial\tilde{r}} {t^\mu_{\phantom{\mu}\nu}}^{\rm ct(2)}+ d\frac{\rho}{\tilde{r}}{t^\mu_{\phantom{\mu}\nu}}^{\rm ct(2)} + \mathcal{L}_\chi {t^\mu_{\phantom{\mu}\nu}}^{\rm ct(2)}+\nonumber\\&&
+\frac{1}{d-2}\frac{1}{\tilde{r}^{d-2}}\Bigg(\nabla^\mu z^\alpha_{\phantom{\alpha}\nu}\nabla_\alpha\rho+\nabla_\nu z^\mu_{\phantom{\mu}\alpha}\nabla^\alpha\rho+z^\mu_{\phantom{\mu}\alpha}\nabla_\nu \nabla^\alpha\rho+z^\alpha_{\phantom{\alpha}\nu}\nabla^\mu \nabla_\alpha\rho -\nonumber\\&&
\qquad\qquad\qquad-2\nabla^\alpha z^\mu_{\phantom{\mu}\nu}\nabla_\alpha\rho -z^\mu_{\phantom{\mu}\nu}\nabla^2\rho -\delta^\mu_{\phantom{\mu}\nu}z^\alpha_{\phantom{\alpha}\beta}\nabla_\alpha\nabla^\beta\rho-\nonumber\\&&
\qquad\qquad\qquad -{\rm Tr}\, z \,\nabla^\mu\nabla_\nu\rho +{\rm Tr}\, z \,\nabla^2 \rho\, \delta^\mu_{\phantom{\mu}\nu}\Bigg) +
\nonumber\\&&
+\frac{2}{\tilde{r}^{d-1}}\left(\nabla^\mu \nabla_\nu \rho - \nabla^2 \rho \,\delta^\mu_{\phantom{\mu}\nu}\right).
\end{eqnarray}
The terms in the last line above are precisely those needed to cancel the divergent terms in the transformation of ${t^\mu_{\phantom{\mu}\nu}}^{\rm bare}$ as in \eqref{tmunubarediffeo}.

We recall that
\begin{equation}
t^\mu_{\phantom{\mu}\nu} ={t^\mu_{\phantom{\mu}\nu}}^{\rm bare}+{t^\mu_{\phantom{\mu}\nu}}^{\rm ct(1)}+{t^\mu_{\phantom{\mu}\nu}}^{\rm ct(2)}+ \mathcal{O}(\nabla^3).
\end{equation}
Therefore combining \eqref{tmunubarediffeo}, \eqref{tmunuct1diffeo} and  \eqref{tmunuct2diffeo} we obtain
\begin{eqnarray}\label{tmunudiffeo}
\tilde{t}{^\mu_{\phantom{\mu}\nu}}  &=&{t^\mu_{\phantom{\mu}\nu}} + \rho \frac{\partial}{\partial\tilde{r}} {t^\mu_{\phantom{\mu}\nu}}+ d\frac{\rho}{\tilde{r}}{t^\mu_{\phantom{\mu}\nu}} + \mathcal{L}_\chi {t^\mu_{\phantom{\mu}\nu}}+\nonumber\\&&
+\frac{1}{d-2}\frac{1}{\tilde{r}^{d-2}}\Bigg(\nabla^\mu z^\alpha_{\phantom{\alpha}\nu}\nabla_\alpha\rho+\nabla_\nu z^\mu_{\phantom{\mu}\alpha}\nabla^\alpha\rho+z^\mu_{\phantom{\mu}\alpha}\nabla_\nu \nabla^\alpha\rho+z^\alpha_{\phantom{\alpha}\nu}\nabla^\mu \nabla_\alpha\rho -\nonumber\\&&
\qquad\qquad\qquad-2\nabla^\alpha z^\mu_{\phantom{\mu}\nu}\nabla_\alpha\rho -z^\mu_{\phantom{\mu}\nu}\nabla^2\rho -\delta^\mu_{\phantom{\mu}\nu}z^\alpha_{\phantom{\alpha}\beta}\nabla_\alpha\nabla^\beta\rho-\nonumber\\&&
\qquad\qquad\qquad -{\rm Tr}\, z \,\nabla^\mu\nabla_\nu\rho +{\rm Tr}\, z \,\nabla^2 \rho\, \delta^\mu_{\phantom{\mu}\nu}\Bigg)+\mathcal{O}(\nabla^3).
\end{eqnarray} 
After using the inverted relation between $z^\mu_{\phantom{\mu}\nu}$ and $t^\mu_{\phantom{\mu}\nu}$ using \eqref{t-z}, and retaining terms up to two derivatives, we obtain \eqref{tdiffeo} for $d>2$ as claimed before when the boundary metric is flat Minkowski space.

\section{Useful identities}\label{uis}
Let us assume that
\begin{equation}
g_{\mu\nu} = g_{(0)\mu\nu}(x) + r^2 g_{(2)\mu\nu}(x) + r^4 g_{(4)\mu\nu}(x) +\mathcal{O}(r^6).
\end{equation}
The exact forms of $g_{(2)\mu\nu}$ and $g_{(4)\mu\nu}$ are not important for the moment. The expansion for the Levi-Civitia connection is:
\begin{equation}
\Gamma^\mu_{\nu\rho}[g] = \Gamma^\mu_{(0)\nu\rho}(x) + r^2 \Gamma^\mu_{(2)\nu\rho}(x) + r^4 \Gamma^\mu_{(4)\nu\rho}(x) + \mathcal{O}(r^6),
\end{equation}
with 
\begin{eqnarray}
 \Gamma^\mu_{(0)\nu\rho} &=&  \Gamma^\mu_{\nu\rho}[g_{(0)}],\nonumber\\
 \Gamma^\mu_{(2)\nu\rho} &=&\frac{1}{2} g^{\mu\sigma}_{(0)}\left(\nabla_{(0)\nu} \, g_{(2)\sigma\rho} +\nabla_{(0)\rho} \, g_{(2)\sigma\nu} -\nabla_{(0)\sigma} \, g_{(2)\nu\rho}  \right), \nonumber\\
 \Gamma^\mu_{(4)\nu\rho} &=&\frac{1}{2} g^{\mu\sigma}_{(0)}\left(\nabla_{(0)\nu} \, g_{(4)\sigma\rho} +\nabla_{(0)\rho} \, g_{(4)\sigma\nu} -\nabla_{(0)\sigma} \, g_{(4)\nu\rho}  \right) -\nonumber\\&&
 -\frac{1}{2} \, g^{\mu\alpha}_{(0)}\, g_{(2)\alpha\beta}\, g^{\beta\sigma}_{(0)}\left(\nabla_{(0)\nu} \, g_{(2)\sigma\rho} +\nabla_{(0)\rho} \, g_{(2)\sigma\nu} -\nabla_{(0)\sigma} \, g_{(2)\nu\rho}  \right),
\end{eqnarray}
where $\nabla_{(0)}$ is the covariant derivative built from $g_{(0)}$. It follows that
\begin{equation}
R^\mu_{\phantom{\mu}\nu\rho\sigma}[g] = R^\mu_{(0)\nu\rho\sigma}(x)+ r^2 R^\mu_{(2)\nu\rho\sigma}(x) + r^4 R^\mu_{(4)\nu\rho\sigma} (x)+ \mathcal{O}(r^6),
\end{equation}
with
\begin{eqnarray}
 R^\mu_{(0)\nu\rho\sigma} &=& R^\mu_{\phantom{\mu}\nu\rho\sigma}[g_{(0)}], \nonumber\\
  R^\mu_{(2)\nu\rho\sigma} &=& \nabla_{(0)\rho} \Gamma^\mu_{(2)\nu\sigma} - \nabla_{(0)\sigma} \Gamma^\mu_{(2)\nu\rho}, \nonumber\\
  R^\mu_{(4)\nu\rho\sigma} &=& \nabla_{(0)\rho} \Gamma^\mu_{(4)\nu\sigma} - \nabla_{(0)\sigma} \Gamma^\mu_{(4)\nu\rho} + \Gamma^\mu_{(2)\rho\alpha}\Gamma^\alpha_{(2)\nu\sigma} -\Gamma^\mu_{(2)\sigma\alpha}\Gamma^\alpha_{(2)\nu\rho}.
\end{eqnarray}
Clearly, the similar coefficients in the expansion for Ricci tensor are,
\begin{equation}
R_{\mu\nu}[g]=R^\alpha_{(0)\mu\alpha\nu}(x)+ r^2 R^\alpha_{(2)\mu\alpha\nu}(x) + r^4 R^\alpha_{(4)\mu\alpha\nu}(x) + \mathcal{O}(r^6),
\end{equation}
and that for the Ricci scalar are:
\begin{eqnarray}
R[g] &=& R_{(0)}+ r^2 \left(R_{(2)\mu\nu} \, g^{\mu\nu}_{(0)}-R_{(0)\mu\nu}\,g^{\mu\alpha}_{(0)}\,g_{(2)\alpha\beta}\, g^{\beta\nu}_{(0)} \right) +\nonumber\\&&
+ r^4 \Big(R_{(4)\mu\nu} \, g^{\mu\nu}_{(0)}-R_{(2)\mu\nu}\,g^{\mu\alpha}_{(0)}\, g_{(2)\alpha\beta}\, g^{\beta\nu}_{(0)} + \nonumber\\ &&\qquad\quad+ R_{(0)\mu\nu} \,g^{\mu\alpha}_{(0)}\left(g_{(2)\alpha\gamma}\,g^{\gamma\delta}_{(0)}\,g_{(2)\delta\beta} - g_{(4)\alpha\beta}\right)g^{(0)\beta\nu}\Big) +
\nonumber\\ &&+ \mathcal{O}(r^6).
\end{eqnarray}

\section{Derivation of highly efficient RG flow equation in conformally flat space}\label{hergCF}

We can solve Einstein's equations perturbatively in $r$ asymptotically when the boundary metric is conformally flat, meaning it is $\eta_{\mu\nu}e^{2\sigma(x)}$ exactly as we have done in Section \ref{calculating-map} when the boundary metric chosen to be flat Minkowski space. However as the conformally flat metric  $\eta_{\mu\nu}e^{2\sigma(x)}$ has a curvature, the subleading term in $g_{\mu\nu}$ is $\mathcal{O}(r^2)$ and not $\mathcal{O}(r^4)$ as in $d=4$. 

To proceed, it is useful to expand around the exact solution which has no term in the Fefferman-Graham expansion beyond $\mathcal{O}(r^4)$. This corresponds to the general case when the \textit{bulk} metric is conformally flat as well. This exact solution is \cite{Skenderis:1999nb}
\begin{equation}\label{es}
g_{\mu\nu}^* = g_{(0)\mu\nu}(x) + r^2 g_{(2)\mu\nu}(x) + r^4 g_{(4)\mu\nu}^*(x),
\end{equation}
with
\begin{equation}
g_{(4)\mu\nu}^* = \frac{1}{4} g_{(2)\mu\alpha}\, g^{\alpha\beta}_{(0)}\,g_{(2)\beta\alpha}.
\end{equation}
The leading term $ g_{(0)\mu\nu}$ being the boundary metric should be
\begin{equation}
 g_{(0)\mu\nu} = e^{2\sigma(x)}\eta_{\mu\nu}.
\end{equation}
The subleading term $g_{(2)\mu\nu}$ takes the form
\begin{equation}
g_{(2)\mu\nu} = - \frac{1}{2}\left(R_{(0)\mu\nu} -\frac{1}{6}R_{(0)} g_{(0)\mu\nu}\right),
\end{equation}
for an arbitrary solution (we have used the notation of Appendix \ref{uis}). More explicitly,
\begin{equation}
g_{(2)\mu\nu} = - (\partial_\mu \sigma) (\partial_\nu\sigma) + \partial_\mu\partial_\nu\sigma + \frac{1}{2}\eta_{\mu\nu} \eta^{\alpha\beta}(\partial_\alpha \sigma) (\partial_\beta \sigma).
\end{equation}

In order to find the general solution, we can write without loss of generality that
\begin{equation}
g_{(4)\mu\nu}(x) = g_{(4)\mu\nu}^*(x) + a_{\mu\nu}(x).
\end{equation}
In this case, the general solution takes the form
\begin{equation}
g_{\mu\nu} = g_{(0)\mu\nu}(x) + r^2 g_{(2)\mu\nu}(x) + r^4 \left(g_{(4)\mu\nu}^*(x) + a_{\mu\nu}(x)\right) + r^6 g_{(6)\mu\nu}(x) + \mathcal{O}(r^8).
\end{equation}
This procedure of perturbing around the exact solution \eqref{es} to find the $g_{(6)\mu\nu}$ for a general solution with a flat boundary metric is advantageous, because on simple dimensional grounds $a_{\mu\nu}$ along with its derivatives can appear only linearly in $g_{(6)\mu\nu}$. Thus we can simply expand \eqref{tensor-equation} linearly about the exact solution \eqref{es} and then truncate the solution at the leading order to find $g_{(6)\mu\nu}$. In turn this will give us the missing term in the general highly efficient RG flow equation in conformally flat space.

In order to satisfy the vector and scalar constraints, namely \eqref{vector-equation} and \eqref{scalar-equation} we require
\begin{eqnarray}\label{aprops}
a_{\mu\nu} \, g^{\mu\nu}_{(0)} = 0, \quad g_{(0)}^{\alpha\beta}\,\nabla_{(0)\alpha}a_{\beta\mu} = 0,
\end{eqnarray}
where $\nabla_{(0)}$ is constructed from $g_{(0)\mu\nu}$. Using the various identities in Appendix \ref{uis}, we obtain
\begin{eqnarray}
g_{(6)\mu\nu} &=& -\frac{1}{12}\left( g_{(0)}^{\alpha\beta}\nabla_\alpha\nabla_\beta a_{\mu\nu}\right) +\frac{2}{3}\left(a_{\mu\alpha}\, g_{(0)}^{\alpha\beta}\, g_{(2)\beta\nu} + g_{(2)\mu\alpha}\, g_{(0)}^{\alpha\beta}\, a_{\beta\nu} \right) -\frac{1}{6}a_{\mu\nu} \,g_{(2)\alpha\beta}\, g_{(0)}^{\alpha\beta} -\nonumber\\&&
- \frac{1}{6}g_{(0)\mu\nu}\, g_{(0)}^{\alpha\gamma}\,g_{(0)}^{\beta\delta}\, a_{\alpha\beta}\, g_{(2)\gamma\delta}.
\end{eqnarray}
Substituting the above in \eqref{t-z4d} and employing the identities in Appendix \ref{uis} once again we can obtain the coefficients of the expansion
\begin{equation}\label{texpansioncf}
t^\mu_{\phantom{\mu}\nu} = t^\mu_{(0)\nu}(x) + r^2 t^\mu_{(2)\nu}(x) + r^2\, \log\,r \, t^\mu_{(2a)\nu}(x) + \mathcal{O}(r^4 \log r)
\end{equation}
of $t^\mu_{\phantom{\mu}\nu}$. Note the term proportional to $\log\, r$ in \eqref{t-z4d} can also be shown to be proportional to the $r^4\,\log \,r$ in the asymptotic expansion of $g_{\mu\nu}$ which vanishes when the boundary metric is conformally flat (as in the exact solution \eqref{es}). Therefore this term does not contribute at the leading order, meaning at order $\log\, r$, and this is why there is no $\log\, r$ term in \eqref{texpansioncf}. However it does contribute at order $r^2\,\log\, r$ giving rise to $t^\mu_{(2a)\nu}$ term in \eqref{texpansioncf}.

The leading term in \eqref{texpansioncf} is
\begin{eqnarray}\label{t0}
t^\mu_{(0)\nu} ={t^\mu_{\phantom{\mu}\nu}}^\infty = {t^{\mu}_{(0)\nu}}^* + 4 g^{\mu\alpha}_{(0)}a_{\alpha\nu},
\end{eqnarray}
with
\begin{eqnarray}\label{t0star}
{t^{\mu}_{(0)\nu}}^*  &=& -g_{(0)}^{\mu\alpha}\,g_{(2)\alpha\beta}\,g_{(0)}^{\beta\gamma}\,g_{(2)\gamma\nu} + g_{(0)}^{\mu\alpha}\,g_{(2)\alpha\nu}\,g_{(0)}^{\beta\gamma}\,g_{(2)\beta\gamma} +\nonumber\\&&+ \frac{1}{2}\delta^\mu_{\phantom{\mu}\nu}\,\left(g_{(0)}^{\alpha\gamma}g_{(2)\alpha\beta}\,g_{(0)}^{\beta\delta}\,g_{(2)\gamma\delta} - \left(g_{(0)}^{\alpha\beta}\,g_{(2)\alpha\beta}\right)^2 \right).
\end{eqnarray}
The term ${t^{\mu}_{(0)\nu}}^*$ equals ${t^\mu_{\phantom{\mu}\nu}}^\infty$ for the exact solution \eqref{es} and is entirely a functional of the boundary metric. The subleading term in \eqref{texpansioncf} is 
\begin{eqnarray}\label{t2}
{t^{\mu}_{(2)\nu}}  &=&  {t^{\mu}_{(2)\nu}}^* +  \hat{t}^{\mu}_{(2)\nu},
\end{eqnarray}
where 
\begin{eqnarray}\label{t2star}
{t^{\mu}_{(2)\nu}}^* &=& g_{(0)}^{\mu\alpha}\,g_{(2)\alpha\beta}\,g_{(0)}^{\beta\gamma}\,g_{(2)\gamma\delta}\,g_{(0)}^{\gamma\delta}g_{(2)\delta\nu} -\frac{1}{2} g_{(0)}^{\mu\alpha}\,g_{(2)\alpha\beta}\,g_{(0)}^{\beta\gamma}\,g_{(2)\gamma\nu}\,g_{(0)}^{\delta\epsilon}\,g_{(2)\delta\epsilon}  -\nonumber\\&&-\frac{1}{2} g_{(0)}^{\mu\alpha}\,g_{(2)\alpha\nu}\,g_{(0)}^{\beta\delta}\,g_{(2)\beta\gamma}\,g_{(0)}^{\gamma\epsilon}\,g_{(2)\delta\epsilon} 
+\nonumber\\&&+ \frac{1}{2}\delta^\mu_{\phantom{\mu}\nu}\Bigg(-g_{(0)}^{\phi\alpha}\,g_{(2)\alpha\beta}\,g_{(0)}^{\beta\gamma}\,g_{(2)\gamma\delta}\,g_{(0)}^{\delta\epsilon}\,g_{(2)\epsilon\phi}
+\nonumber\\&&\qquad\qquad+\left(g_{(0)}^{\alpha\gamma}\,g_{(2)\alpha\beta}\,g_{(0)}^{\beta\delta}\,g_{(2)\gamma\delta}\right)\left( g_{(0)}^{\epsilon\phi}\,g_{(2)\epsilon\phi}\right) \Bigg),
\end{eqnarray}
is an explicit functional of the boundary metric and 
\begin{eqnarray}\label{t2hat}
\hat{t}^{\mu}_{(2)\nu} = -g^{\alpha\beta}_{(0)}\,\nabla_{(0)\alpha}\nabla_{(0)\beta}\,(g^{\mu\rho}_{(0)}\,a_{\rho\nu}) + 4 g_{(0)}^{\mu\alpha}\,a_{\alpha\beta}\,g_{(0)}^{\beta\gamma}\, g_{(2)\gamma\nu} + 6\delta^\mu_{\phantom{\mu}\nu} \, g_{(0)}^{\alpha\gamma}\,a_{\alpha\beta}g_{(0)}^{\beta\delta}\,g_{(2)\gamma\delta}.
\end{eqnarray}
Also
\begin{eqnarray}\label{t2a}
t^\mu_{(2a)\nu} &=& \text{$r^2$ term in the expansion of}\nonumber\\&&
\Bigg(\frac{1}{8}R^\mu_{\phantom{\mu}\alpha\nu\beta}[g]R^{\alpha\beta}[g]- \frac{1}{48}\nabla^\mu\nabla_\nu R[g]+ \frac{1}{16}\nabla^2 R^{\mu}_{\phantom{\mu}\nu}[g]-\frac{1}{24}R[g]R^{\mu}_{\phantom{\mu}\nu}[g]+\nonumber\\&&\qquad\qquad+\left(\frac{1}{96}R^2[g]-\frac{1}{32}R_{\alpha\beta}[g]R^{\alpha\beta}[g]-\frac{1}{96}\nabla^2 R[g]\right)\delta^{\mu}_{\phantom{\mu}\nu}\Bigg),
\end{eqnarray}
which is thus an explicit functional of the boundary metric as well. This can be readily obtained from the compact but explicit expansions of $\Gamma^\mu_{\nu\rho}[g]$, $R^\mu_{\phantom{\mu}\nu\rho\sigma}[g]$ etc. obtained in Appendix \ref{uis}. The exact form of this is a lengthy expression which is not too illuminating, so we skip giving a more detailed expression here.

Using \eqref{texpansioncf}, \eqref{t0}, \eqref{t0star}, \eqref{t2}, \eqref{t2star}, \eqref{t2hat} and  \eqref{t2a}, we can readily invert the relation between $a_{\mu\nu}$ and $t^\mu_{\phantom{\mu}\nu}$. We obtain,
\begin{eqnarray}
\eta^{\mu\alpha}e^{-2\sigma}a_{\alpha\nu} &=& \frac{1}{4}\left(t^\mu_{\phantom{\mu}\nu} -{t^\mu_{(0)\nu}}^* \right) - r^2 \Big({t^{\mu}_{(2)\nu}}^* + \nonumber\\&&\qquad\qquad\qquad\qquad\qquad
+\frac{1}{4}\Big(-g^{\alpha\beta}_{(0)}\,\nabla_{(0)\alpha}\nabla_{(0)\beta}\,\left(t^\mu_{\phantom{\mu}\nu} -{t^\mu_{(0)\nu}}^* \right) 
+ \nonumber\\&&\qquad\qquad\qquad\qquad\qquad\qquad
+4 \left(t^\mu_{\phantom{\mu}\beta} -{t^\mu_{(0)\beta}}^* \right)\,g_{(0)}^{\beta\gamma}\, g_{(2)\gamma\nu} 
+ \nonumber\\&&\qquad\qquad\qquad\qquad\qquad\qquad+ 6\delta^\mu_{\phantom{\mu}\nu} \, \left(t^\gamma_{\phantom{\gamma}\beta} -{t^\gamma_{(0)\beta}}^* \right)g_{(0)}^{\beta\delta}\,g_{(2)\gamma\delta}\Big)\Big) \nonumber\\&&- r^2 \,\log \, r\, t^\mu_{(2a)\nu} + \mathcal{O}(r^4\, \log \, r).
\end{eqnarray}
Differentiating \eqref{texpansioncf} and using the above inverted relation we finally obtain the highly efficient RG flow equation for $t^\mu_{\phantom{\mu}\nu}$, which is as follows:
\begin{eqnarray}
\frac{\partial}{\partial r} t^\mu_{\phantom{\mu}\nu} &=& r \Big(2{t^{\mu}_{(2)\nu}}^* + \nonumber\\&&\qquad\qquad
+\frac{1}{2}\Big(-g^{\alpha\beta}_{(0)}\,\nabla_{(0)\alpha}\nabla_{(0)\beta}\,\left(t^\mu_{\phantom{\mu}\nu} -{t^\mu_{(0)\nu}}^* \right) 
+ \nonumber\\&&\qquad\qquad\qquad
+4 \left(t^\mu_{\phantom{\mu}\beta} -{t^\mu_{(0)\beta}}^* \right)\,g_{(0)}^{\beta\gamma}\, g_{(2)\gamma\nu} 
+ \nonumber\\&&\qquad\qquad\qquad+ 6\delta^\mu_{\phantom{\mu}\nu} \, \left(t^\gamma_{\phantom{\gamma}\beta} -{t^\gamma_{(0)\beta}}^* \right)g_{(0)}^{\beta\delta}\,g_{(2)\gamma\delta}\Big)+ \nonumber\\&&\qquad+ t^\mu_{(2a)\nu}\Big) + 2 r\, \log \, r \, t^\mu_{(2a)\nu} +\mathcal{O}(r^3 \log \, r).
\end{eqnarray}
Replacing $r$ by $\Lambda^{-1}$ above we obtain \eqref{t-rg-example-CF}. We have also replaced $g_{(0)}^{\mu\nu}$ by its more explicit form $e^{-2\sigma}\eta^{\mu\nu}$, and changed notations by replacing $\nabla_{(0)}$ by $\nabla^{(\sigma)}$.

\section{Derivation of transformations under bulk diffeomorphisms in UV expansion}\label{diffeotransformderivationUV}
Here we proceed by assuming that the boundary metric is conformally flat, meaning it is $\eta_{\mu\nu}e^{2\sigma(x)}$, and that $d=4$. We would be interested to derive the transformation of $t^\mu_{\phantom{\mu}\nu}$ up to $\mathcal{O}(r^4 \log \, r)$. Furthermore, we will restrict ourselves to two derivatives.

Assuming minimalist counterterms (those required for cancellations of UV divergences only), only the first two counterterms and the log counterterm can contribute to the transformation of $t^\mu_{\phantom{\mu}\nu}$ up to $\mathcal{O}(r^4 \log \, r)$. We can thus proceed exactly as in Appendix \ref{diffeotransformderivation} and obtain the transformation by expanding \eqref{tmunudiffeo} up to required orders. Note all the results in Appendix \ref{diffeotransformderivation} are generally valid for any arbitrary boundary metric, except for\eqref{tdiffeo} where we have imposed that the boundary metric is flat Minkowski space. On top of this we need to add the transformation of the log counterterm. However this $log$ counterterm only contributes at sixth order in derivatives, so we can ignore its effect if we restrict ourselves to two derivatives.

Doing the UV (asymptotic) expansion of $z^\mu_{\phantom{\mu}\nu}$, $\rho$; $\chi^\mu$, $g_{\mu\nu}$ and $\nabla$ etc. in \eqref{tmunudiffeo} we get
\begin{eqnarray}\label{tmunudiffeoCF}
\tilde{t}{^\mu_{\phantom{\mu}\nu}}  &=&{t^\mu_{\phantom{\mu}\nu}} + \Bigg(\tilde{r}^2 \,\delta\sigma\, \frac{1}{2}\Big(-g^{\alpha\beta}_{(0)}\,\nabla_{(0)\alpha}\nabla_{(0)\beta} t^\mu_{\phantom{\mu}\nu}
+4 t^\mu_{\phantom{\mu}\beta} \,g_{(0)}^{\beta\gamma}\, g_{(2)\gamma\nu} 
+  6\delta^\mu_{\phantom{\mu}\nu} \, t^\gamma_{\phantom{\gamma}\beta}\,g_{(0)}^{\beta\delta}\,g_{(2)\gamma\delta}\Big)\nonumber\\&& + 4\,\delta\sigma \,{t^\mu_{\phantom{\mu}\nu}} - \frac{1}{2}\tilde{r}^2 \mathcal{L}_{e^{-2\sigma}\eta^{\alpha\beta}\partial_\beta\delta\sigma}\,\, {t^\mu_{\phantom{\mu}\nu}}+\nonumber\\&&
+\frac{\tilde{r}^2 e^{-2\sigma}}{2}\Bigg(\eta^{\mu\beta}\partial_\beta t^\alpha_{\phantom{\alpha}\nu}\,\partial_\alpha\delta\sigma+\partial_\nu t^\mu_{\phantom{\mu}\alpha}\,\partial_\beta\delta\sigma\,\eta^{\alpha\beta}+t^\mu_{\phantom{\mu}\alpha}\,\partial_\nu \partial_\beta\delta\sigma\,\eta^{\alpha\beta}+t^\alpha_{\phantom{\alpha}\nu}\,\partial_\beta \partial_\alpha\delta\sigma\,\eta^{\beta\mu}-\nonumber\\&&
\qquad\qquad\qquad-2\partial_\alpha t^\mu_{\phantom{\mu}\nu}\,\partial_\beta\delta\sigma\,\eta^{\alpha\beta} -t^\mu_{\phantom{\mu}\nu}\,\partial_\alpha\partial_\beta\delta\sigma\,\eta^{\alpha\beta} -\delta^\mu_{\phantom{\mu}\nu}t^\alpha_{\phantom{\alpha}\beta}\,\partial_\alpha\partial_\gamma\delta\sigma\,\eta^{\beta\gamma}\Bigg)+\nonumber\\&&+\mathcal{O}\left(\tilde{r}^4\right)\Bigg)+\mathcal{O}(\partial^3),
\end{eqnarray} 
up to fourth order in derivatives and given orders in the asymptotic expansion. Replacing $r$ by $\Lambda^{-1}$ above we obtain \eqref{tmunudiffeoCFLambda}. We have also replaced $g_{(0)}^{\mu\nu}$ by its more explicit form $e^{-2\sigma}\eta^{\mu\nu}$, and changed notations by replacing $\nabla_{(0)}$ by $\nabla^{(\sigma)}$.

% Please avoid comments such as "For a review'', "For some examples",
% "and references therein" or move them in the text. In general,
% please leave only references in the bibliography and move all
% accessory text in footnotes.

% Also, please have only one work for each \bibitem.

\bibliographystyle{utcaps}
\bibliography{HolographyAsRGFLOW-refs}

\end{document}